\documentclass[sigconf]{acmart}
\usepackage{amsmath}
\usepackage{array}
\usepackage{makecell}
\usepackage{hyperref}
\hypersetup{
	colorlinks=true,
	linkcolor=blue,
	filecolor=blue,      
	urlcolor=blue,
	citecolor=cyan,
}
\usepackage{algorithm}
\usepackage{algpseudocode}
\usepackage{multirow}
\usepackage{booktabs}
\usepackage{stfloats}
\usepackage{wrapfig} 
\AtBeginDocument{%
  }

\setcopyright{acmlicensed}
\copyrightyear{2024}
\acmYear{2024}
\acmDOI{10.1145/3658644.3690221}

\acmConference[CCS '24] {Proceedings of the 2024 ACM SIGSAC Conference on Computer and Communications Security}{October 14--18, 2024}{Salt Lake City, UT, USA.}
\acmBooktitle{Proceedings of the 2024 ACM SIGSAC Conference on Computer and Communications Security (CCS '24), October 14--18, 2024, Salt Lake City, UT, USA}
\acmISBN{979-8-4007-0636-3/24/10}

\begin{document}

\title{TREC: APT Tactic / Technique Recognition via Few-Shot Provenance Subgraph Learning}


\author{Mingqi Lv}
\affiliation{%
  \institution{College of Computer Science and Technology, Zhejiang University of Technology}
  \city{Hangzhou}
  \country{China}}
\email{mingqilv@zjut.edu.cn}

\author{Hongzhe Gao}
\affiliation{%
	\institution{College of Computer Science and Technology, Zhejiang University of Technology}
	\city{Hangzhou}
	\country{China}}
\email{211122120114@zjut.edu.cn}

\author{Xuebo Qiu}
\affiliation{%
	\institution{College of Computer Science and Technology, Zhejiang University of Technology}
	\city{Hangzhou}
	\country{China}}
\email{qiuxuebo@zjut.edu.cn}

\author{Tieming Chen}
\affiliation{%
	\institution{College of Computer Science and Technology, Zhejiang University of Technology}
	\city{Hangzhou}
	\country{China}}
\email{tmchen@zjut.edu.cn}

\author{Tiantian Zhu}
\affiliation{%
	\institution{College of Computer Science and Technology, Zhejiang University of Technology}
	\city{Hangzhou}
	\country{China}}
\email{ttzhu@zjut.edu.cn}

\author{Jinyin Chen}
\affiliation{%
	\institution{College of Computer Science and Technology, Zhejiang University of Technology}
	\city{Hangzhou}
	\country{China}}
\email{chenjinyin@zjut.edu.cn}

\author{Shouling Ji}
\affiliation{%
	\institution{College of Computer Science and Technology, Zhejiang University}
	\city{Hangzhou}
	\country{China}}
\email{sji@zju.edu.cn}

%
%
%
%
%

\renewcommand{\shortauthors}{Mingqi Lv et al.}

\begin{abstract}
 APT (Advanced Persistent Threat) with the characteristics of persistence, stealth, and diversity is one of the greatest threats against cyber-infrastructure. As a countermeasure, existing studies leverage provenance graphs to capture the complex relations between system entities in a host for effective APT detection. In addition to detecting single attack events as most existing work does, understanding the tactics / techniques (e.g., Kill-Chain, ATT\&CK) applied to organize and accomplish the APT attack campaign is also important for security operations. Existing studies try to manually design a set of rules to map low-level system events to high-level APT tactics / techniques. However, the rule based methods are coarse-grained and lack generalization ability. Thus, they can only recognize APT tactics and have difficulty in identifying APT techniques. They also cannot adapt to mutant behaviors of existing APT tactics / techniques.

In this paper, we propose TREC, the first attempt to recognize APT tactics / techniques from provenance graphs by exploiting deep learning techniques. To address the “needle in a haystack” problem, TREC segments small and compact subgraphs covering individual APT technique instances from a large provenance graph based on a malicious node detection model and a subgraph sampling algorithm. To address the “training sample scarcity” problem, TREC trains the APT tactic / technique recognition model in a few-shot learning manner by adopting a Siamese neural network. We evaluate TREC based on a customized dataset collected and made public by our team. The experiment results show that TREC significantly outperforms state-of-the-art systems in APT tactic recognition and TREC can also effectively identify APT techniques.
\end{abstract}




\maketitle

\section{Introduction}
The long-standing war between defenders and attackers in computing systems keeps evolving. The network attacks under the theme of APT (Advanced Persistent Threat) have grown more sophisticated and destructive. Although security infrastructure like EDR (Endpoint Detection and Response) and IDS (Intrusion Detection System) have been broadly deployed, APT attacks can still easily penetrate the defenses and cause severe damages\cite{r1}.

As compared with the simple hit-and-run cyber-attack strategies, APT attacks are \textit{persistent} and \textit{stealthy}. APT attackers could hide within a host undetected for a long period of time and leverage a variety of APT tactics / techniques to bypass the defend systems step by step. To effectively detect the complex APT attacks, the detection systems should capture the long-term casual and contextual information of the fine-grained behaviors of the hosts. Therefore, recent studies utilize provenance data collected from low-level kernel audit logs for APT detection \cite{r2,r3}. The provenance data can be represented as an acyclic graph (called a \textit{provenance graph}), where the nodes stand for system entities (e.g., processes, files, sockets) and the edges stand for system events (e.g., fork, write, open). The provenance graphs can organize the contextual information of each system entity in a structured way.

\begin{figure}[!htbp]
	\centering
	\includegraphics[width=0.45\textwidth]{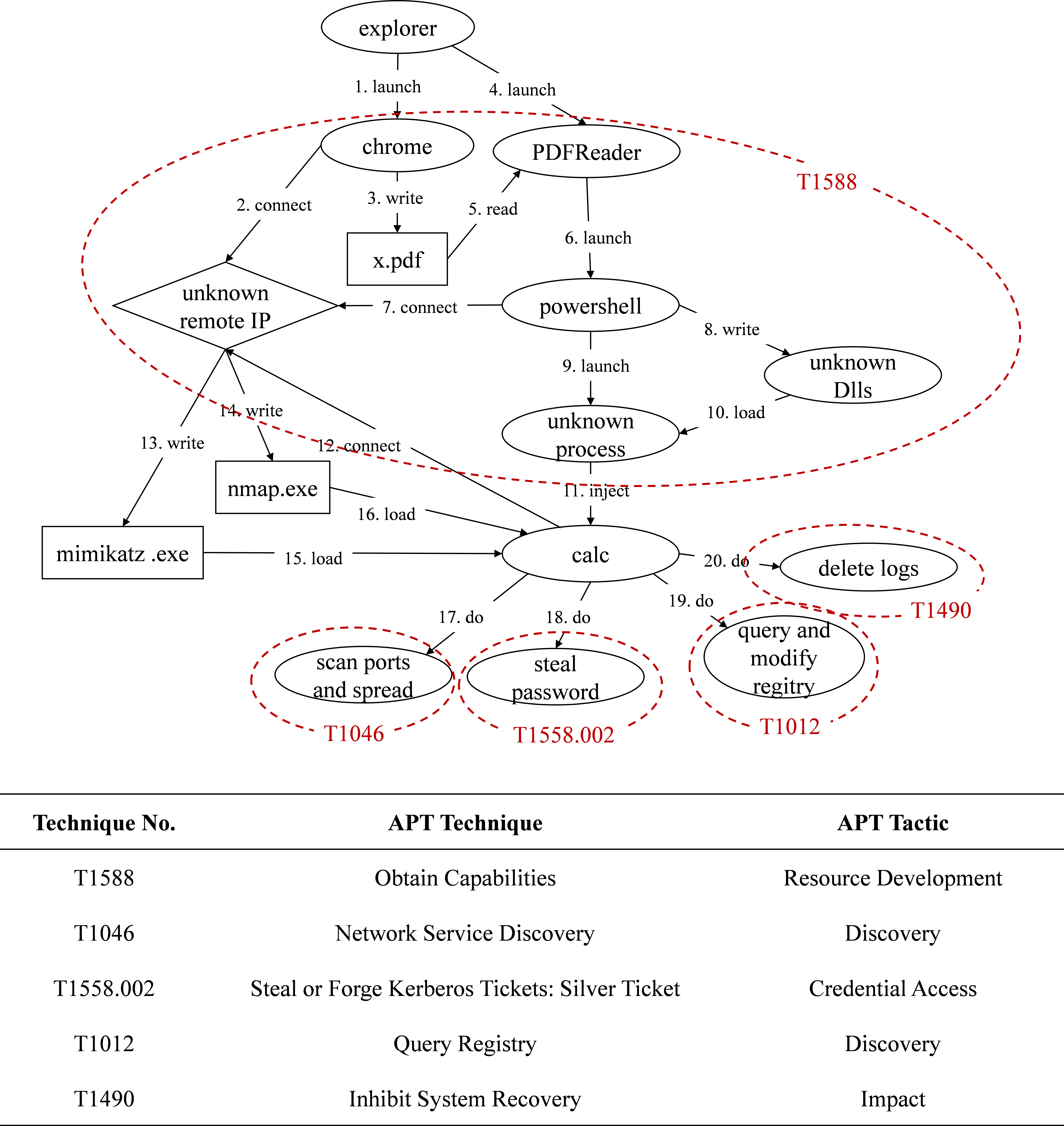}
	\caption{An illustration of APT attack analysis based on APT tactics / techniques.}
	\label{fig:1}
	\vspace{-0.5cm}
\end{figure}

Due to the persistence of APT attacks, the provenance graphs can be extremely large and complex. Even a single host can generate tens of millions of system events per day \cite{r4}. Hence, the accurate detection and tracing of APT attacks in raw provenance graphs is like searching a needle in a haystack. Aiming at this challenge, existing studies try to map the system events in raw provenance graphs to APT tactics / techniques based on attack knowledge such as Kill-Chain \cite{r5} and ATT\&CK \cite{r6}, and then discover real APT attacks by following certain APT tactic / technique patterns. For example, Figure \ref{fig:1} shows a snapshot of the provenance graph from an APT attack campaign, which is composed of five APT techniques. First, the victim user connects to an unknown remote IP via Chrome and downloads a PDF file containing malicious codes. Once the PDF file is opened, the macro in it would be executed to create a PowerShell process, which creates another unknown process to execute the malicious operations by injecting the malicious codes into the calc process (\textit{T1588}). Then, the infected calc process would scan ports and spread the malicious script (\textit{T1046}), steal password by using the Mimikatz tool (\textit{T1588.002}), query and modify registry (\textit{T1012}), and finally deletes backup files (\textit{T1490}). The five APT techniques and the corresponding APT tactics are listed in the bottom of Figure \ref{fig:1}. With the APT technique chain (\textit{T1588}, \textit{T1046}, \textit{T1588.002}, \textit{T1012}, and \textit{T1490}), the APT attacks could be well traced and understood out of huge number of provenance graph nodes.

The majority of existing work detects APT attacks based on binary classification, e.g., classifying a provenance graph node as benign or malicious \cite{r9,r13,r29}. Without pinpointing the detailed APT tactics / techniques, it would make cybersecurity analysts suffer from discovering true APT attacks among an enormous number of alerts \cite{r27}. Aiming at this problem, several existing studies try to recognize APT tactics / techniques in raw provenance graphs using rule based methods \cite{r7,r8,r9}, where a predefined rule database is applied to map each system event to a specific APT tactic / technique. Some examples of rules are shown in Table \ref{tab:1} (\textit{S}, \textit{P}, and \textit{F} stand for socket, process, and file). However, the rule based methods have the following drawbacks. First, the human defined rules are arbitrary and can only represent the superficial patterns of APT tactics / techniques, which can be easily confused with specific benign activities. Therefore, the rule based methods would usually cause a high degree of false alarms. Second, the rules are often defined based on hard knowledge (e.g., whitelists, blacklists), which is difficult to generalize to new or mutant attacks. Thus, the rule based methods are prone to recognition errors even with only slight mutation in attack behaviors of APT tactic / technique instances.

\begin{table*}[t]
	\renewcommand{\arraystretch}{2}
	\centering
	\caption{Examples of APT tactic / technique recognition rules.} 
	\label{tab:1} 
	\resizebox{0.95\textwidth}{!}{
		\begin{tabular}{cccc}
			\hline
			\textbf{APT Stage} & \textbf{APT Technique} & \textbf{Detection Rule} & \textbf{Description} \\
			\hline
			Initial Access & Untrusted remote services &\makecell[l] {$S.ip \notin \{ Trusted\_IP\_Addresses\}$} & \makecell[l]{A socket reading data from\\ untrusted IP address} \\ [0.5em]
			Execution & Shell execution & \makecell[l]{$\begin{array}{l} P.command \in \{ Sensitive\_Commands\} 
					\wedge \exists Initial\_Access(P'):\\Dependence(P,P') > thres \end{array}$} & \makecell[l]{A process executing sensitive \\commands and correlating with \\an initial access process}\\ [0.5em]
			Exfiltration & Sensitive read & \makecell[l]{$\begin{array}{l} F.path \in \{ Sensitive\_Files\} 
					\wedge \exists Initial\_Access(P'):\\Dependence(F,P') > thres\end{array}$} & \makecell[l]{A sensitive file being \\read and  correlating with\\ an initial access process} \\[0.5em]
			\hline
		\end{tabular}
	}
	
\end{table*}

On the other hand, learning based methods (i.e., machine learning and deep learning) are capable of discovering latent patterns from raw data in an automatic way. Therefore, by training on a large corpus of labeled samples, they are expected to accurately recognize APT tactic / technique instances with no need of human labor. For example, we could label the APT tactics / techniques of a large number of system events from provenance graphs, and then train a model to map the behavior features of a new system event to the corresponding APT tactic / technique. However, the learning based APT tactic / technique recognition methods have not been widely applied in practice due to the following challenges.

\textbf{The scarcity of training samples (C1)}: APT attacks are extremely rare events, and thus it is almost impossible to collect a large number of real APT attack samples. What’s worse, there are hundreds and thousands of APT techniques. For example, the current version of ATT\&CK Matrix involves 227 APT techniques, some of which have more sub-techniques. Thus, an individual APT technique would have even less samples.

\textbf{The difficulty of fine-grained sample labeling (C2)}: APT attacks are stealthy, so the malicious activities of APT attacks are usually hidden in tens of thousands of system entities in a provenance graph. In addition, the APT attacks often conceal their malicious activities in normal system entities (e.g., process injection, file contamination). As a result, it is like looking for needles in a haystack to perform a binary benign / malicious labeling on all nodes in a provenance graph, let alone performing a multi-class APT tactic / technique labeling.

\textbf{The confusion of sample segmentation (C3)}: A malicious operation involves a number of system entities, and an APT technique instance usually involves multiple malicious operations. However, the malicious operations and the associated system entities of an APT technique instance can be mixed up with massive other benign system entities and system events, and thus it is extremely difficult to determine the boundary of the subgraphs representing the APT technique instances.

To address these challenges, this paper proposes TREC, an intelligent APT tactic / technique recognition method based on few-shot provenance subgraph learning. For challenge C2, TREC discovers potential malicious nodes (called NOIs, nodes of interest) in the provenance graphs based on an unsupervised learning strategy without the need of labeling. Specifically, it learns normal system activity patterns from benign samples and detects NOIs whose activity greatly deviates from normal system activity patterns. For challenge C3, TREC uses a subgraph sampling algorithm to identify correlated NOIs to form the subgraphs covering individual APT technique instances. For challenge C1, TREC adopts a Siamese neural network based few-shot learning model to recognize APT tactics / techniques from subgraphs, by training on only a small number of samples from each individual APT techniques. In summary, the main contributions of this paper are as follows.

(1) We propose TREC, which is the first attempt for recognizing APT tactics / techniques based on provenance data by exploiting deep learning techniques, with no need of the manually defined rule sets. It tries to address the problems of sample scarcity and labeling difficulty by combining techniques including anomaly detection, subgraph sampling, and few-shot learning.

(2) We propose an APT technique subgraph sampling algorithm, which can segment subgraphs covering individual APT technique instances from large provenance graphs with massive benign nodes by exploiting unsupervised learning and correlative graph node analysis.

(3) We propose a few-shot APT tactic / technique recognition model by combining Siamese neural network and GNN (Graph Neural Network). It transforms the APT tactic / technique recognition task from a classification problem into a contrastive analysis problem to reduce the demand for labeled samples of each APT tactic / technique.

(4) We collected and made public a dataset\footnote{https://www.kellect.org/\#/kellect-4-aptdataset} through a series of APT technique simulations. To the best of our knowledge, it is currently the publicly available dataset that involves the most types of APT techniques. We conducted extensive experiments upon this dataset. The experiment results show that TREC outperforms state-of-the-art rule based APT tactic / technique recognition systems.

\section{Preliminary}
\subsection{Definitions}
\begin{definition}
	(\textit{Provenance Graph}): A provenance graph is defined as $PG=(V,E)$,where the node set $V$ represents all system entities and the edge set $E$ represents all system events. Each edge $e=(u,v,p)$ denotes that system entity $u$ (called subject) performs operation $p$ on system entity $v$ (called object).
\end{definition}
\begin{definition}
	(\textit{Technique Subgraph}): A technique subgraph $TSG$ is a subgraph segmented from a provenance graph. It contains system events belonging to the same APT technique instance. Note that it is usually impossible to construct perfect technique subgraphs in practice. In this paper, we try to find system entities that are strongly correlated with each other to form compact subgraphs as potential technique subgraphs.
	
	Take the provenance graph shown in Figure \ref{fig:1} as an example, the third edge denotes a system event that the subject (process “chrome”) performs a “write” operation on the object (file “x.pdf”). The oval with label “\textit{T1588}” is a technique subgraph covering an instance of the APT technique “\textit{Obtain Capabilities}”. Note that the ovals with labels \textit{T1046}, \textit{T1558.002}, \textit{T1012}, and \textit{T1490} are not raw provenance graph nodes, but the abstract of a subgraph for presentation convenience.
\end{definition}
\begin{definition}
	(\textit{NOI}): A NOI (Node of Interest) is a system entity that performs malicious activity in an APT attack campaign. Note that due to the propagation effect of malicious labels (i.e., the malicious labels would be propagated to all the nodes that are connected to the nodes that are initially labeled as malicious, resulting in too many malicious nodes)\cite{r10}, it might be ambiguous whether a node is a NOI. In this paper, we detect NOIs as nodes whose behaviors have great difference with normal nodes.
\end{definition}

\subsection{Threat Model}
Our threat model is similar to the previous work on provenance-based APT detection \cite{r27,r32,r40}. First, we assume the integrity of the operation system and audit modules, and thus the provenance data are considered credible. Second, we assume that the attackers can use any tactics and techniques to conduct APT attacks, but their malicious behaviors are captured by the audit modules.

\begin{figure}[h]
	\centering
	\includegraphics[width=0.48\textwidth]{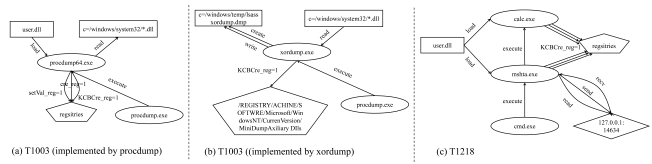}
	\caption{APT technique instances: (a) T1003 implemented by script “procdump”; (b) T1003 implemented by script “xordump”; (c) T1218.}
	\label{fig:2}
	\vspace{-0.3cm}
\end{figure}

In addition, we also assume that APT techniques have distinct patterns, which can be captured by statistical or semantic features \cite{r40}. Specifically, the instances of the same APT technique would leave similar patterns, while different APT techniques would exhibit different patterns. To illustrate the assumption, we show three APT technique instances in Figure \ref{fig:2}. The first one and the second one are of the same APT technique with different implementations. It can be observed that they have similar patterns (e.g., creating new processes, reading DLL files, and modifying registry keys sequentially). The third instance is of a different APT technique, and thus it show different patterns (e.g., creating new processes and accessing to remote server). Note that each APT technique uniquely corresponds to an APT tactic, so the APT tactic can be directly obtained based on the APT technique recognition result.

\subsection{Data Collection}
To collect large-scale provenance data to cover the APT techniques in ATT\&CK, we used Atomic Red Team\footnote{https://github.com/redcanaryco/atomic-red-team}, an automatic security test tool provided by Red Canary, to simulate APT attack activities in a semi-automatic way, and collected provenance dataset by using KELLECT\footnote{https://www.kellect.org/}, an efficient kernel level provenance data collection tool developed by our research team.

Specifically, each script provided by the Atomic Red Team covers a specific APT technique in ATT\&CK. To emulate APT techniques, we set up a secured virtual environment in Windows by using VMware\footnote{https://www.vmware.com/} , and then use PowerShell tool to fully execute the scripts by granting it the system privileges. To make the collected dataset more in line with real application scenarios, we also conduct redundant activities (e.g., Internet surfing, office document editing, file transferring, software installation, and video playback) during the data collection phase.

Each collected system event is a triad (subject, operation, object), and is modeled as two nodes connecting by an edge. Each system entity has a unique identifier, and thus the provenance graph can be created by merging the “object” of a system event and the “subject” of another system event that refer to the same system entity into a node. In particular, the provenance graphs can capture the causality relationships between system entities and facilitate the reasoning over system entities that are temporally distant. Therefore, the provenance graphs are useful in navigating through the stealthy and persistent APT attacks.

\subsection{System Architecture}
\begin{figure}[!htbp]
	\centering
	\includegraphics[width=0.45\textwidth]{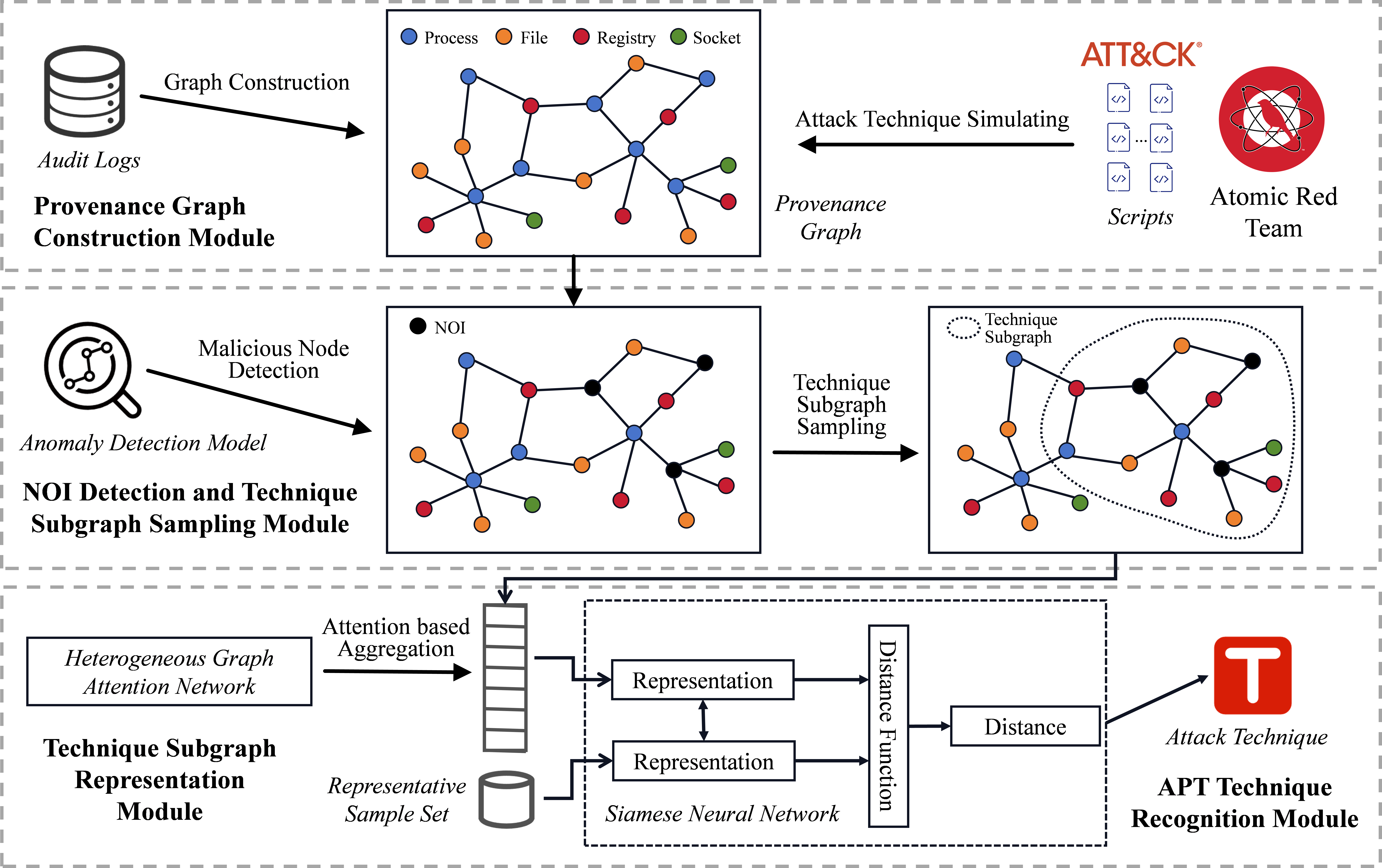}
	\caption{The system architecture of TREC.}
	\label{fig:3}
	\vspace{-0.5cm}
\end{figure}

The architecture of TREC is shown in Figure \ref{fig:3}, consisting of four modules.

\textbf{The provenance graph construction module} (Section 2.3): It continuously collects kernel level system events and arranges these system events to construct a provenance graph.

\textbf{The NOI detection and subgraph sampling module} (Section 4.1): It firstly applies an anomaly detection model to detect potential malicious nodes (i.e., NOIs) in the provenance graph with no need of labeled malicious node samples for training. Then, since NOIs might contain false alarms and an APT technique usually involves many malicious nodes, it adopts a subgraph sampling algorithm to find correlated NOIs and segment them from the provenance graph to form a technique subgraph, which covers the system events of an individual APT technique instance.

\textbf{The technique subgraph representation module} (Section 4.2): Given a technique subgraph $TSG$, it uses heterogeneous graph embedding technique to encode $TSG$ into a feature vector. Since the nodes in a technique subgraph have significantly different degree of impact on the representation of certain APT techniques (a large proportion of nodes are even benign), it applies a hierarchical graph attention mechanism to aggregate the information of nodes in a technique subgraph with different weights.

\textbf{The APT technique recognition module} (Section 4.3): It trains an APT technique recognition model based on deep learning technique. Since it is extremely difficult to collect adequate training samples for each APT technique, it applies a Siamese neural network based few-shot learning strategy to train the model, which requires only a few of training samples for each APT technique to achieve a stable generalization ability.

\section{Methodology}
\subsection{NOI Detection and Subgraph Sampling}
An APT technique usually involves a series of malicious activities used to achieve a specific goal, which can be a step of an APT attack campaign. As shown in Figure \ref{fig:1}, the technique subgraph inside the oval with label “\textit{T1588}” denotes the APT technique “\textit{Obtain Capabilities}”, where the attackers try to obtain capabilities to support their operations by downloading malicious tools\footnote{http://attack.mitre.org/techniques/T1588/}.

However, in practice, the technique subgraphs covering individual APT technique instances are hidden in a huge number of benign system entities, and we have to separate them from the complete provenance graph before further processing. To address this problem, we propose a subgraph sampling method that works in two steps. First, it detects potential malicious nodes (i.e., NOIs) in the original provenance graph. Second, it applies a heuristic subgraph sampling algorithm to segment the technique subgraphs by leveraging the NOIs.

\subsubsection{NOI detection}
The purpose of NOI detection is to quickly discover potential malicious nodes in the provenance graph. The most straightforward methods include designing a set of detection rules to filter out benign nodes or training a classification model to recognize malicious nodes. However, both the two methods have great limitation in practice. First, the detection rule based method lacks generalization ability and cannot detect new or mutant malicious activities. Second, the classification model based method requires a large corpus of fine-grained node-level labeled training samples, which are infeasible as discussed in Section 1. To this end, we design a NOI detection model that is trained in an unsupervised style. The model works in three steps, i.e., feature initialization, feature learning, and NOI detection.

\textbf{Feature initialization:} The features of a node should reflect its behavior pattern in the provenance graph. According to the findings in \cite{r13}, the malicious nodes usually have different interaction pattern with its neighboring nodes as compared with the benign nodes. Therefore, we extract features for each node as follows. First, we categorize the nodes in a provenance graph into four types (including process, file, registry, and socket), and the interactions between different types of nodes can form 21 types of edges (as shown in Table \ref{tab:2}). Then, for each node $v_i$ in the provenance graph, we create a 42-dimensional vector $e_i= [a_1, a_2, …, a_{21}, a_{22}, a_{23}, …, a_{42}]$ as the initial features of $v_i$, where $a_i(1 \le i \le 21)$ is the number of incoming edges of $v_i$  with the $i$-th type and $a_j(22 \le j \le 42)$ is the number of outgoing edges of $v_i$ with the ($j$-21)-th type.For example, the initial feature vector of the node “powershell” in Figure 1 is [1, 0, 0, 0, 0, 0, 0, 0, 0, 0, 0, 0, 0, 0, 0, 0, 0, 0, 0, 0, 0, 1, 0, 0, 1, 0, 0, 0, 0, 0, 0, 0, 0, 0, 0, 0, 0, 0, 1, 0, 0, 0].

\begin{table}[t]
	\renewcommand{\arraystretch}{1.5}
	\centering
	\caption{The summary of edge types.} 
	\label{tab:2} 
	\resizebox{\linewidth}{!}{
		\begin{tabular}{|c|c|c|}
			\hline
			\textbf{Node Type Pairs} & \textbf{Edge Type ID} & \textbf{Edge Types}\\
			\hline
			process - process & $ET_1$ &  \makecell[l]{launch}\\
			\hline
			process - file & $ET_2$ & \makecell[l]{create, read, write, close, delete, enum}\\
			\hline
			process - registry & $ET_3$ & \makecell[l]{open, query, enumerate, modify, close,\\ delete}\\
			\hline
			process - socket & $ET_4$ & \makecell[l]{send, receive, retransmit, copy,\\ connect, disconnect, accept, reconnect}\\
			\hline
		\end{tabular}
	}
	\vspace{-0.5em}
\end{table}

\textbf{Feature learning}: The initial feature vector of a node can only reflect its first-order interaction patterns, which cannot capture long-range causal correlations, and thus cannot adapt to the stealthy and persistent APT attacks. To learn higher-order interaction patterns, we use a GNN (Graph Neural Network) to aggregate information from each node’s ancestors. Most GNNs have to be trained with supervised signals (e.g., downstream tasks), but the malicious / benign labels of nodes are unavailable in this task. Aiming at this problem, we train the GNN in a self-supervised style by leveraging supervised signals inherently implied in the provenance graph. Specifically, we use node types as the supervised signals, and train the GNN with a downstream task of node type classification (i.e., classifying a node into process, file, registry, or socket).

The GNN can be trained through multiple layers. In the $t$-th layer, the node feature matrix $\textbf{E}^{(t)}$ updated according to Equation \ref{equ:1}, where $G$ is the topology of the provenance graph (usually represented by an adjacent matrix) and $\textbf{W}^{(t-1)}$ is a trainable parameter matrix. Note that more layers (i.e., a larger value of $t$) indicate that the GNN can aggregate information from more remote nodes to learn higher-order interaction patterns. Here, $GNN(\cdot)$ is a GNN based encoder, which can be instantiated by any GNN models (e.g., GCN\cite{r14}, GAT\cite{r15}, GIN\cite{r16}, etc.).

\begin{equation}
	\label{equ:1}
	{\textbf{E}^{(t)}} = GNN\left( {{\textbf{W}^{(t - 1)}},G,{\textbf{E}^{(t - 1)}}} \right)
\end{equation}

We stack $T$ GNN layers on the provenance graph $G$ to learn the final feature matrix $\textbf{E}^{(T)}$, which is then input into a softmax classifier to make node type classification. After training, $GNN(\cdot)$ can be treated as a feature extractor. Specifically, given the initial feature matrix $\textbf{E}^{(0)}$ of a provenance graph, we input $\textbf{E}^{(0)}$ into the trained encoder $GNN(\cdot)$, which would output the final feature matrix $\textbf{E}^{(T)}$  of the provenance graph.

\textbf{NOI detection:} After obtaining the final feature vector for every node in the provenance graph, we treat the NOI detection as an outlier mining problem. Specifically, we apply iForest\cite{r17}, which has been proven to be an effective outlier mining algorithm, to detect outliers in all “process” nodes in the provenance graph. These outliers are treated as NOIs.

\subsubsection{Subgraph sampling}
The purpose of subgraph sampling is to separate the technique subgraphs covering individual APT technique instances from the original provenance graph. To achieve this purpose, we design a subgraph sampling algorithm by leveraging the detected NOIs as clues through three steps, i.e., seed node selecting, graph searching, and subgraph sampling. The core idea of the subgraph sampling algorithm is that the malicious activities of an APT technique instance would usually have strong correlations with each other, resulting in a clustering effect of the NOIs in the provenance graph.

\textbf{Seed node selecting:} In network science, a hub node is a node that has a significant higher degree of connectivity in the graph. The hub nodes usually play a crucial role in graph analysis. In provenance graph, a hub node is also more likely to correlate with malicious activities\cite{r13}. For example, a malicious node related to the APT technique “\textit{Exfiltration over C2 Channel}” usually has a large number of “connect” and “send” edges to remote IP nodes. A malicious node related to the “\textit{Living off the Land}” APT technique usually have a large number of “launch” edges for executing sensitive instructions (e.g., the “powershell” node). Thus, we select the NOI with the highest degree of connectivity as the seed node.

\textbf{Graph searching:} Starting from a seed node $v_k$, we firstly perform a DFS (Depth-First Search), going forward or backward, to find out whether we can reach other NOIs in $\lambda$ hops, where $\lambda$ is a pre-defined length threshold of search paths. If no NOI could be found in a $\lambda$-hop search path, we stop expanding the search along this path. Otherwise, we recursively start the DFSs from each found NOI to expand the search paths.

\textbf{Subgraph sampling:} When the graph searching process from seed node $v_k$ stops, a technique subgraph $TSG_k = (SV_k, SE_k)$ will be created where $SV_k$ is the node set containing all the visited nodes during the graph searching process (including both NOIs and benign nodes) and $SE_k$ is the corresponding edge set.

\subsection{Technique Subgraph Representation}
To classify the technique subgraphs into different APT tactics / techniques, they have to be embedded into feature vectors by aggregating the feature vectors of all the involved nodes. However, it is a challenging task due to the following reasons. First, the technique subgraphs involve multiple types of nodes and edges, which should be taken into account when learning the representations. Second, due to the diversity of APT techniques, the technique subgraphs could have great difference in the sizes. In our dataset, the largest technique subgraph contains 20841 nodes and the smallest one contains 436 nodes. Third, each technique subgraph contains a large portion of benign nodes, and thus different nodes in a technique subgraph should have great differences in the impact on the final representation.

Aiming at these challenges, we design a technique subgraph embedding model by extending HAN (Heterogeneous Graph Attention Network)\cite{r18}. Formally, the purpose of technique subgraph embedding is to learn a function that maps each technique subgraph $TSG = (SV, SE)$ to a $d$-dimensional feature vector $\textbf{h}_{TSG} (d \ll |SV|)$, where $SV$ and $SE$ are node set and edge set, respectively.The feature vector $\textbf{h}_{TSG}$ should be capable of preserving the key semantics in $TSG$.The core idea of this model is twofold. First, since provenance graphs are heterogeneous, we apply meta-path scheme\cite{r19} to help the model to learn node / edge type aware graph traversal paths. Second, we use attention mechanism to force the technique subgraph representations to focus on more important parts of the subgraph. To this end, we perform technique subgraph embedding through two steps, i.e., meta-path defining and heterogeneous technique subgraph embedding.

\textbf{Meta-path defining:} A meta-path is defined as a sequence of node / edge types in the form of ${A_1}\xrightarrow{{{R_1}}}{A_2}\xrightarrow{{{R_2}}} \cdots \xrightarrow{{{R_l}}}{A_{l + 1}}$ where $A_k$ denotes a node type and $R_k$ denotes an edge type.Meta-paths could provide a structured way to specify and capture the relationships between different types of nodes in the heterogeneous graph. Based on the node and edge types in Table \ref{tab:2}, we define a variety of meta-paths over different types of system entities from different views, as summarized in Table \ref{tab:3}. For example, $MP_2$ means that processes accessing to the same file have similar semantics. We denote the set of all defined meta-paths as $MPS$.

\begin{table}[t]
	\renewcommand{\arraystretch}{1.4}
	\centering
	\caption{The summary of meta-paths.} 
	\label{tab:3} 
	\resizebox{0.35\textwidth}{!}{
		\begin{tabular}{cc}
			\hline
			\textbf{ID} & \textbf{Meta-paths}\\
			\hline
			$MP_1$ &  \makecell[l]{$process\xrightarrow{{E{T_1}}}process$}\\
			
			$MP_2$ &  \makecell[l]{$process\xrightarrow{{E{T_2}}}file\xrightarrow{{ET_2^{ - 1}}}process$} \\
			
			$MP_3$ &  \makecell[l]{$process\xrightarrow{{E{T_3}}}registry\xrightarrow{{ET_3^{ - 1}}}process$}\\
			
			$MP_4$ &  \makecell[l]{$process\xrightarrow{{E{T_4}}}socket\xrightarrow{{ET_4^{ - 1}}}process$}\\
			\hline
		\end{tabular}
	}
\end{table}

\textbf{Heterogeneous technique subgraph embedding:} The hierarchical attention mechanism used for the heterogeneous technique subgraph embedding includes a node-level attention, a path-level attention, and a graph-level attention.

In the node-level attention, for each node $v_i$ in the technique subgraph, we try to measure the importance of each node in its neighbors. Specifically, given the node $v_i$ and a meta-path $MP_j$, we travel $MP_j$ starting from $v_i$ to form a set of nodes (denoted as $N_i^j$), which connect with $v_i$ via $MP_j$ (including $v_i$ itself). Then, we apply self-attention\cite{r20} to learn the attention weight of each node $v_k$ in $N_i^j$ on $v_i$ (denoted as $\alpha_{ki}^j$) based on Equation \ref{equ:3},where $e_k$ and $e_i$ are the initial feature vectors of nodes $v_k$ and $v_i$ in Section 3.2.1, and $att_{node}(\cdot)$ is the self-attention function.Finally, the feature vector of $v_i$ is updated by aggregating the feature vectors of all nodes in $N_i^j$ with the corresponding attention weights based on Equation \ref{equ:4}, where $h_i^j$ is the updated feature vector of $v_i$ under meta-path $MP_j$, and $\sigma(\cdot)$ is an activation function. In summary, the node-level attention is able to capture the meta-path aware semantic information for each node.

\begin{equation}
	\label{equ:3}
	\alpha _{ki}^j = softmax\left( {at{t_{node}}({\textbf{e}_k},{\textbf{e}_i})} \right)
\end{equation}

\begin{equation}
	\label{equ:4}
	\textbf{h}_i^j = \sigma \left( {\sum\nolimits_{{v_k} \in N_i^j} {\alpha _{ki}^j}  \cdot {\textbf{e}_k}} \right)
\end{equation}

In the path-level attention, for each node $v_i$ in the technique subgraph, after obtaining the updated feature vectors of $v_i$ for all meta-paths (denoted as $\{h_i^j\}_{j=1}^{|MPS|}$),we try to measure the importance of each meta-path for $v_i$. Specifically, given the node $v_i$ and a meta-path $MP_j$, we compute the attention weight of $MP_j$ on $v_i$ (denoted as $\beta_i^j$) based on Equation \ref{equ:5}, where $\textbf{W}_1$ is a trainable parameter matrix, $\textbf{q}$ and $\textbf{b}$ are trainable parameter vectors. After obtaining the attention weight for each meta-path in \textit{MPS}, we can fuse the meta-path aware feature vectors to obtain the final feature vector of $v_i$ (denoted as $\textbf{h}_i$) based on Equation \ref{equ:6}. In summary, the path-level attention is able to capture the aggregated contextual semantic information for each node.

\begin{equation}
	\label{equ:5}
	\beta _i^j = softmax\left( {{\textbf{q}^T} \cdot \tanh ({\textbf{W}_1} \cdot \textbf{h}_i^j + \textbf{b})} \right)
\end{equation}

\begin{equation}
	\label{equ:6}
	{\textbf{h}_i} = \sum\nolimits_{M{P_j} \in MPS} {\beta _i^j \cdot \textbf{h}_i^j} 
\end{equation}

In the graph-level attention, we try to generate one embedding for the technique subgraph \textit{TSG} based on the final feature vectors of all nodes in \textit{SV}. Since different nodes in \textit{TSG} play different roles and show different importance, we also use an attention function to compute the attention weight of each node in \textit{SV} before the embedding fusion. Specifically, we firstly compute a global context vector \textbf{c} by simply averaging the node feature vectors followed by a nonlinear transformation based on Equation \ref{equ:7},where $\textbf{W}_2$ is a trainable parameter matrix. Then, we compute the attention weight for each node $v_i$ (denoted as $\gamma_i$) that is aware of the global context vector \textbf{c} based on Equation \ref{equ:8}, which is the inner product between \textbf{c} and the final feature vector of $v_i$. The intuition is that nodes similar to the global context should receive higher attention weights. Finally, we compute the technique subgraph embedding (denoted as \textbf{h}) as the weighted sum of node feature vectors based on Equation \ref{equ:9}. The overall process of the proposed heterogeneous technique subgraph embedding is shown in Algorithm \ref{alg:1}.

\begin{equation}
	\label{equ:7}
	\textbf{c} = \tanh \left( {\frac{1}{{|SV|}}{\textbf{W}_2}\sum\nolimits_{{v_i} \in SV} {{\textbf{h}_i}} } \right)
\end{equation}

\begin{equation}
	\label{equ:8}
	{\gamma _i} = softmax\left( {\textbf{h}_i^T \cdot \textbf{c}} \right)
\end{equation}

\begin{equation}
	\label{equ:9}
	\textbf{h} = \sum\nolimits_{{v_i} \in SV} {{\gamma _i} \cdot {\textbf{h}_i}}
\end{equation}

\begin{algorithm}
	\caption{The Heterogeneous Technique Subgraph Embedding Algorithm}\label{alg:1}
	\begin{algorithmic}[1]
		\Require The technique subgraph, $TSG = (SV, SE)$ \label{alg:input1}
		\Require The initial feature matrix of $SV$, $E$ ($e_i$ denotes the feature matrix of $v_i$) \label{alg:input2}
		\Require The set of meta-paths, $MPS$ \label{alg:input3}
		\Ensure The technique subgraph embedding \label{alg:output}
		\For{$v_i \in SV$}
		\For{$MP_j \in MPS$}
		\State Find the meta-path based neighbors, $N_i^j$
		\For{$v_k \in N_i^j$}
		\State Calculate the node-level attention weight, $\alpha_{ki}^j$
		\EndFor
		\State Aggregate the node-level feature vectors, $\textbf{h}_i^j = \sigma\left(\sum_{v_k \in N_i^j} \alpha_{ki}^j \cdot \textbf{e}_k\right)$
		\State Calculate the path-level attention weight, $\beta_{i}^j$
		\EndFor
		\State Aggregate the path-level feature vectors, $\textbf{h}_i = \sum_{MP_j \in MPS} \beta_{i}^j \cdot \textbf{h}_i^j$
		\EndFor
		\For{$v_i \in SV$}
		\State Calculate the graph-level attention weight, $\gamma_{i}$
		\EndFor
		\State Aggregate the graph-level feature vectors, $\textbf{h} = \sum_{v_i \in SV} \gamma_{i} \cdot \textbf{h}_i$
		\State \textbf{Return} $\textbf{h}$
	\end{algorithmic}
\end{algorithm}

\subsection{APT Technique Recognition}
After obtaining the embedding of a technique subgraph that covers an APT technique instance, the most intuitive way for APT technique recognition is to build a classification model, which takes the technique subgraph embedding as input and outputs the APT technique label of it. However, this intuitive method could be infeasible in practice due to challenge C1. First, since we cannot collect adequate training samples for all APT techniques in practice, it is almost impossible to train a multi-class classification model with high generalization ability to effectively capture and distinguish the diverse patterns of different APT techniques. Second, the APT techniques are constantly evolving. When we need to consider a new APT technique, the entire classification model has to be retrained to adapt to the new label space.

Aiming at these problems, we adopt a Siamese neural network based few-shot learning method to build the APT technique recognition model. The core idea is to convert the APT technique recognition from a classification task into a contrastive analysis task. Specifically, Siamese neural network is based on the coupling framework established by twin neural networks, which share the same structure and parameters \cite{r21}. The Siamese neural network takes two samples as inputs, which are embedded into a common feature space based on the twin neural networks. Then, they are joined by a distance function to measure the distance between the feature vectors of the two inputs. A well trained Siamese neural network would maximize that of different labels and minimize the distance of the representations of the same label.

The Siamese neural network can well address the above mentioned two problems. First, it trains to determine the distance between pairs of technique subgraph samples rather than model the explicit pattern of each APT technique, and thus it does not require so many training samples for each APT technique. Second, the Siamese neural network does not have to be retrained when considering a new APT technique. Instead, it can recognize the new APT technique with even only one training sample by measuring the distance between a new sample and this training sample in a one-shot learning style.

\begin{figure}[h]
	\centering
	\includegraphics[width=0.48\textwidth]{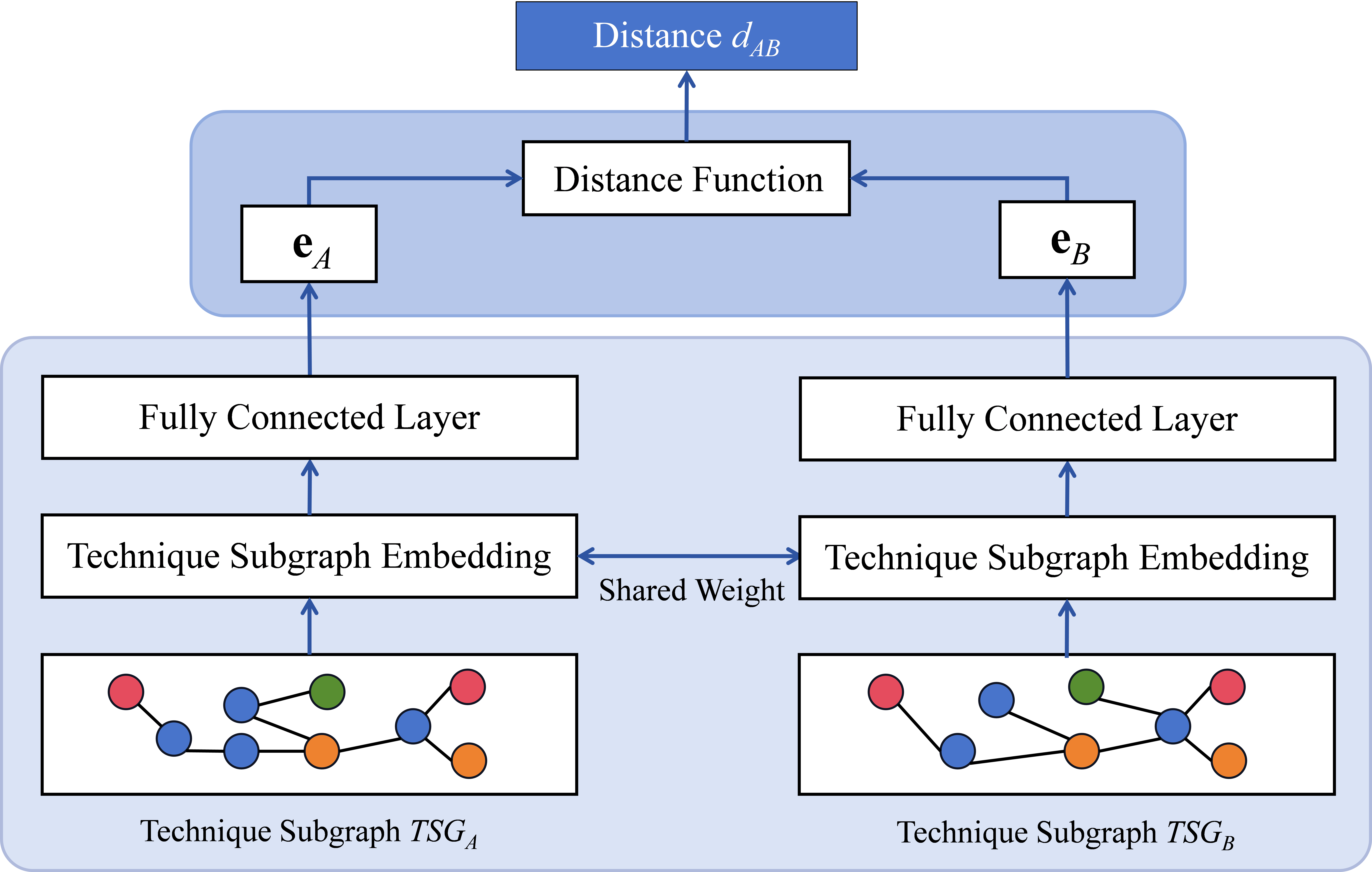}
	\caption{The architecture of the proposed Siamese neural network.}
	\label{fig:6}
	
\end{figure}

\textbf{Model architecture:} As shown in Figure \ref{fig:6}, the proposed Siamese neural network consists of three layers, i.e., the input layer, the encoding layer, and the interactive layer. The input layer and the encoding layer have two branches. First, each branch of the input layer accepts a technique subgraph (i.e., $TSG_A$) representing an individual APT technique instance as input. Second, each branch of the encoding layer applies the technique subgraph embedding model proposed in Section 3.2 to transform the technique subgraph into a feature vector (i.e., $\textbf{h}_A$), and then stacks a fully connected layer to further process the feature vector to obtain a final embedding (i.e., $\textbf{e}_A$).Third, the interactive layer uses distance function (e.g., Euclidean distance, cosine similarity, etc.) to calculate the distance between the final embeddings of the two branches (i.e., $d_{AB}$).

\textbf{Model training:} To train the Siamese neural network, we adopt the anchor-positive-negative triplet selection strategy to construct the training set. Specifically, for each available technique subgraph sample, we treat it as an anchor sample (denoted as $s_a$), and randomly select a positive sample from the same APT technique (denoted as $s_p$) and a negative sample from a different APT technique (denoted as $s_n$). The purpose of this selection strategy is to balance the number of positive and negative pairs. Then, we could split the triplet into two sample pairs $x_1 = (s_a, s_p)$ and $x_2 = (s_a, s_n)$, and adopt them to train the Siamese neural network with the aim to minimize the contrastive loss function as Equation \ref{equ:10}, where $d$ is the distance between two input samples and $m$ is the margin. The loss function encourages the representations of samples from the same APT technique to be close and representations of samples from different APT techniques to be far apart.

\begin{equation}
	\label{equ:10}
	\begin{gathered}
		loss\left( {{s_a},{s_p}} \right) = {d^2} \hfill \\
		loss\left( {{s_a},{s_n}} \right) = \max \left( {0,m - d} \right) \hfill \\ 
	\end{gathered} 
\end{equation}

\textbf{Model inference:} Given a new technique subgraph sample $s_k$, we perform APT technique recognition as a matching task. First, for each APT technique $C_i$, we pick a representative sample for $C_i$ (i.e., $s_i$). For example, we can choose the medoid of an APT technique as the representative sample\cite{r22}. Second, we input $s_k$ and $s_i$ into the two branches of the trained Siamese neural network to obtain two final embeddings (i.e., $e_k$ and $e_i$). Finally, we calculate the distance between $e_k$ and $e_i$ (denoted as $d_{ki}$). After that, we could classify $s_k$ to the APT technique with the smallest distance. We could also set a lower-bound distance threshold to discover outlier samples that are not belong to any known APT technique.

\section{Experiment}
\subsection{Experiment Setup}
\subsubsection{Dataset}
The existing public provenance datasets are not suitable for the evaluation of APT tactic / technique recognition methods due to the following reasons. First, the existing datasets do not involve a rich set of APT techniques. Second, the existing databases have not provided APT tactic / technique labels. In contrary, they only provide graph-level binary labels (e.g., StreamSpot) or node-level binary labels (e.g., ATLAS, DARPA TC) to indicate a provenance graph or a system entity is malicious or not.

Therefore, we collect our own dataset according to Section 2.3. We use scripts provided by Atomic Red Team to simulate a variety of APT techniques. Note that each script can simulate a specific APT technique, so the provenance graph sample collected via one execution of a script always contains only one technique subgraph. We label the boundary and APT technique of the technique subgraph in each collected provenance graph sample. Finally, the dataset contains 473 provenance graph samples, which involve 26 APT techniques and 45 APT sub-techniques from 9 APT tactics. Each APT technique / sub-technique has an average of only 18 / 10 samples, so it is a typical few-shot learning problem.

The average number of nodes and NOIs of the provenance graph samples are 8957 and 4, respectively. The detailed list of the involved APT tactics / techniques is shown in Appendix A. We have publicly released our dataset at \url{https://www.kellect.org/#/kellect-4-aptdataset}.

\subsubsection{Evaluation strategies}
There are three tasks in TREC to be evaluated, i.e., the NOI detection task, the technique subgraph sampling task, and the APT tactic / technique recognition task. For the NOI detection task, we apply a Leave-Malicious-Out evaluation strategy, i.e., the NOI detection model is trained on a dataset with all benign nodes, and tested on a dataset with malicious nodes and the same number of randomly sampled benign nodes. Accuracy, Precision, Recall, F1 Score, FAR (False Alarm Rate), and AUC (Area Under Curve) are used as evaluation metrics. Here, Precision, Recall, and F1 Score are calculated for malicious nodes. FAR refers to the ratio of benign nodes that are mistakenly detected as malicious. For the technique subgraph sampling task, we evaluate its performance by comparing the sampled technique subgraphs and ground-truth technique subgraphs based on several customized evaluation metrics defined in Section 4.3.

For the APT tactic / technique recognition task, we evaluate its performance under three conditions, i.e., \textit{True\_Graph}, \textit{Sampled\_Graph}, and \textit{Raw\_Graph}, where \textit{True\_Graph} uses the ground-truth technique subgraphs as input to the APT tactic / technique recognition model, \textit{Sampled\_Graph} uses the sampled technique subgraphs based on the algorithm in Section 3.1 as input, and \textit{Raw\_Graph} uses the raw provenance graph samples without sampling as input. Here, \textit{True\_Graph} can specifically evaluate the recognition ability of the model, while \textit{Sampled\_Graph} can evaluate the overall performance of TREC. TacACC, TechACC, and SubACC are used as evaluation metrics, which stand for accuracies of APT tactic recognition, APT technique recognition, and APT sub-technique recognition, respectively.
\begin{table}[t]
	\renewcommand{\arraystretch}{1.4}
	\caption{Comparison of Models on \textit{True\_Graph} and \textit{Sampled\_Graph}}
	\centering
	\label{tab:4} 
	\resizebox{0.48\textwidth}{!}{
		\begin{tabular}{c|ccc|ccc}
			\hline
			& \multicolumn{3}{c|}{\textit{True\_Graph}} & \multicolumn{3}{c}{\textit{Sampled\_Graph}} \\
			Models& TacACC & TechACC & SubACC & TacACC & TechACC & SubACC \\ 
			\hline
			GCN\_C & 0.690 & 0.430 & 0.215 & 0.380 & 0.290 & 0.097 \\
			GCN\_S & 0.704 & 0.473 & 0.441 & 0.473 & 0.309 & 0.215 \\
			HAN\_C & 0.859 & 0.538 & 0.258 & 0.803 & 0.326 & 0.118 \\
			TREC   & \textbf{0.875} & \textbf{0.797} & \textbf{0.766} & \textbf{0.813} & \textbf{0.703} & \textbf{0.672} \\
			\hline
		\end{tabular}
	}
\end{table}

\subsection{Experiment 1: Comparison Experiment}
In the first experiment, we compare TREC with the following GNN based models.

(1) GCN\_C: It is a homogeneous graph neural network based classification model. Specifically, it first creates a homogeneous graph with the same topology for each technique subgraph. Second, it performs node embedding based on a three-layered GCN and summarizes the node embeddings into a graph embedding based on a simple average pooling operation. Finally, it trains a classifier based on the graph embeddings as the APT tactic / technique recognition model.

(2) GCN\_S: It is a homogeneous graph neural network based similarity matching model. It uses the same method with GCN\_C to obtain technique subgraph embeddings, but trains the recognition model based on the Siamese neural network.

(3) HAN\_C: It is a heterogeneous graph neural network based classification model. Specifically, it uses the same method with TREC to obtain technique subgraph embeddings (i.e., using HAN to obtain node embeddings and using a context based attention mechanism to obtain graph embedding), but trains the recognition model as a classifier.

The experiment results are shown in Table 4. First, all the models have much higher TacACC than their TechACC and SubACC. The first reason is that the number of APT techniques / sub-techniques far exceeds the number of APT tactics. The second reason is that different APT tactics usually have great differences in behavior patterns, while different APT techniques / sub-techniques may have trivial differences in provenance graphs. Second, TREC outperforms GCN\_S and HAN\_C outperforms GCN\_C. It indicates that heterogeneous graphs can better capture the semantics of provenance graphs than homogeneous graphs do. The types of system entities and system events are essential information for APT tactic / technique recognition, and thus GCN\_C and GCN\_S would inevitably produce misclassifications when the technique subgraphs of different APT techniques share similar topological structure. Third, TREC outperforms HAN\_C and GCN\_S outperforms GCN\_C. It shows that matching models are more suitable for our task than classification models. In our task, there are a large number of APT techniques with very limited samples for each one, and thus directly training a classification model would easily become overfitting. Overall, TREC has the best performance on all evaluation metrics by encoding the technique subgraphs based on heterogeneous graph neural networks and training the model in a few-shot learning manner.

In the second experiment, we compare TREC with two state-of-the-art methods, i.e., HOLMES \cite{r36} and APTShield \cite{r38}. First, the two methods are based on heuristic rules designed by referring to attack knowledge Kill-Chain and ATT\&CK, but they only provide the overall design concepts and a part of concrete rules. Second, the rules of the two methods cover most APT tactics in Kill-Chain and ATT\&CK, but cannot distinguish the finer-grained APT techniques. Therefore, we re-implement the two methods and compare the APT tactic recognition performance with them. The details of re-implementation are shown in Appendix B and the overall design concepts of the two methods are as follows.

(1) HOLMES: It categorizes APT tactics into seven classes based on Kill-Chain. It designs a number of rules (called TTP specifications) to map low-level audit logs to an intermediate layer of sensitive activities (e.g., shell execution, sensitive file reading, C\&C commands, etc.), and then it recognizes APT tactics through the correlation analysis of the sensitive activities.

(2) APTShield: It categorizes APT tactics into ten classes based on ATT\&CK. It designs a large number of suspicious labels to characterize various suspicious system entities and system events. The initial suspicious labels are propagated and aggregated based on a set of transfer rules considering the data flow and control flow in different APT tactics. Then, it recognizes APT tactics by referring to the aggregated suspicious labels.

The experiment results are shown in Table \ref{tab:5}. Note that we compare with HOLMES by testing the Kill-Chain based APT tactic recognition and compare with APTShield by testing the ATT\&CK based APT tactic recognition.

TREC significantly outperforms HOLMES by up to 40\%. The reasons are as follows. First, the rule design of HOLMES heavily relies on whitelists (e.g., trusted IP addresses, super-user groups) and blacklists (e.g., sensitive files, sensitive commands). Thus, the recognition performance depends on the integrity of the whitelists and blacklists. Second, an APT tactic can be implemented by a variety of APT techniques. However, the rules of HOLMES are static and their number is limited, so they cannot cover all the APT techniques simulated in the experiment. Third, the rules of HOLMES consider causal correlations in a single path. For example, rule represented by Equation \ref{equ:11} is used to recognize the \textit{Establish a Foothold} APT tactic, which indicates process \textit{P} has executed command line utilities and the shortest path between \textit{P} and the initial compromise point is less than \textit{path\_thres}. Obviously, it cannot capture the multi-head contextual correlations in the neighbors of \textit{P}, so these rules are less effective in recognizing more stealthy APT attacks.

\begin{equation}
	\label{equ:11}
	\begin{gathered}
		F.path \in \{ Command\_Line\_Utilities\}  \hfill \\
		\wedge \exists Initial\_Compromise(P'):\\path\_factor(P',P) <  = path\_thres \hfill \\ 
	\end{gathered} 
\end{equation}

APTShield improves the recognition performance of HOLMES by incorporating more suspicious system entity states and a fine-grained suspicious state transfer algorithm for each APT tactic in ATT\&CK. The recognition accuracy of TREC is 13\% higher than APTShield. The reasons are as follows. First, APTShield also heavily relies on whitelists and blacklists. However, the sensitive system entities defined for different APT tactics may overlap. For example, if a process \textit{P} operates a sensitive file \textit{F} that is defined in the blacklists of two different APT tactics, \textit{P} would be assigned with multiple states that confuse the recognition rules. Second, the items defined in a blacklist can also interact with benign system entities. For example, a benign process reading a sensitive file would also trigger a rule for recognizing the \textit{Data Exfiltration} APT tactic in APTShield, resulting in a false positive. Third, the suspicious state transfer algorithm would amplify the negative influence of poorly defined rules.

In addition, TREC can better adapt to real application environment than HOLMES and APTShield do. First, the two rule based methods can only recognize APT tactic and cannot recognize APT technique, since it is too difficult to manually design fine-grained rules to distinguish the sophisticated APT techniques. On the other hand, TREC is totally data-driven, and it can automatically learn latent and generalizable patterns hidden in provenance graphs that are difficult to discover by human experts. Second, the heuristic rules are designed for experiment environment (e.g., the blacklists are specifically defined for the dataset), so their recognition performance would further degrade when applying in real application environment. On the other hand, TREC can adapt to new environment by retraining or fine-tuning using new training samples.

\begin{table}[t]
	\renewcommand{\arraystretch}{1.4}
	\caption{The comparison with state-of-the-art rule based methods.}
	\centering
	\label{tab:5} 
	\resizebox{0.48\textwidth}{!}{
		\begin{tabular}{c|cc|cc}
			\hline
			& \multicolumn{2}{c|}{Kill-Chain Tactic Recognition} & \multicolumn{2}{c}{ATT\&CK Tactic Recognition} \\
			& HOLMES & TREC & APTShield & TREC  \\ 
			\hline
			ACC & 0.535 & 0.930 & 0.774 & 0.875 \\
			\hline
		\end{tabular}
	}
\end{table}

\subsection{Experiment 2: Parameter Tuning Experiment}
In this section, we investigate the impact of the two key parameters in TREC, i.e., the dimension of technique subgraph embeddings $d$ (Section 3.2) and the length threshold of search paths for technique subgraph sampling $\lambda$ (Section 3.1.2).

\begin{figure}[h]
	\centering
	\includegraphics[width=0.48\textwidth]{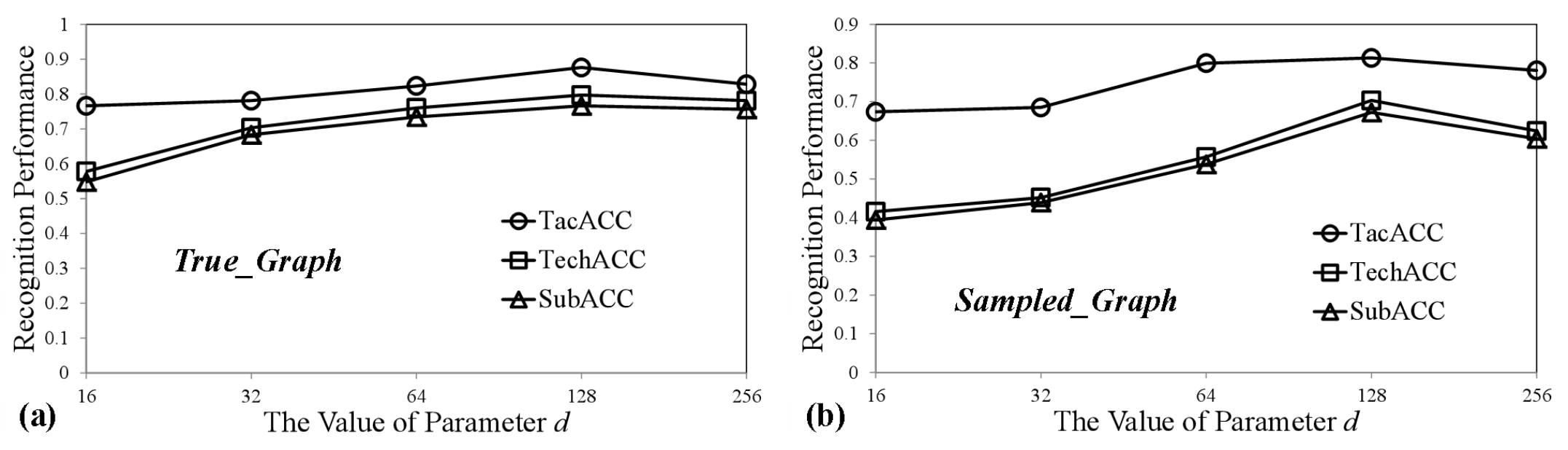}
	\caption{The tuning of parameter $d$: (a) the tuning on \textit{True\_Graph}; (b) the tuning on \textit{Sampled\_Graph}.}
	\label{fig:7}
\end{figure}

In the first experiment, we tune parameter d by monitoring the APT tactic / technique recognition performance. The experiment results are shown in Figure \ref{fig:7}. By adjusting $d$ in the range of [16, 256], the performance generally exhibits an increasing trend, which indicates that it is difficult to encode enough topological and semantic information of the technique subgraphs when $d$ is too small. However, when increasing $d$ from 128 to 256, the increasing trend becomes stable or even shows a slight decrease. It suggests that a too large $d$ could potentially lead to overfitting by adding noises to the embedding vectors. To this end, we set $d = 128$ in the following experiment.

In the second experiment, we tune parameter $\lambda$ by monitoring the technique subgraph sampling performance, which is evaluated using four metrics, i.e., Precision, Coverage, TPR (True Positive Rate), and FAR (False Alarm Rate). Here, Precision and Coverage are used to measure the degree of overlap between the ground-truth technique subgraphs and sampled technique subgraphs. Specifically, given a sampled technique subgraph $sTSG_k$ and its corresponding ground-truth technique subgraph $gTSG_k$, we denote the set of ground-truth NOIs in $gTSG_k$, as $G\_NS_k$ and the set of detected NOIs in $sTSG_k$ as $D\_NS_k$. The Precision and Coverage are calculated based on Equation \ref{equ:12} and \ref{equ:13}, where $N$ is the number of samples. In addition, we define a sampled technique subgraph $TSG$ as correct if the Precision and Coverage of $TSG$ are both greater than 0.8. Then, given the set of ground-truth technique subgraphs $gTGS$ and the set of sampled technique subgraphs $sTGS$, in which the set of correct sampled technique subgraphs is $cTGS$, the TPR and FAR are calculated based on Equation \ref{equ:14} and \ref{equ:15}.

\begin{equation}
	\label{equ:12}
	Precision = \frac{1}{N}\sum\nolimits_{k = 1}^N {\frac{{|D\_N{S_k}| \cap |G\_N{S_k}|}}{{|D\_N{S_k}|}}}
\end{equation}

\begin{equation}
	\label{equ:13}
	Coverage = \frac{1}{N}\sum\nolimits_{k = 1}^N {\frac{{|D\_N{S_k}| \cap |G\_N{S_k}|}}{{|G\_N{S_k}|}}}
\end{equation}

\begin{equation}
	\label{equ:14}
	TPR = \frac{{|cTGS|}}{{|gTGS|}}
\end{equation}

\begin{equation}
	\label{equ:15}
	FAR = \frac{{|sTGS - cTGS|}}{{|sTGS|}}
\end{equation}

\begin{figure}[h]
	\centering
	\includegraphics[width=0.48\textwidth]{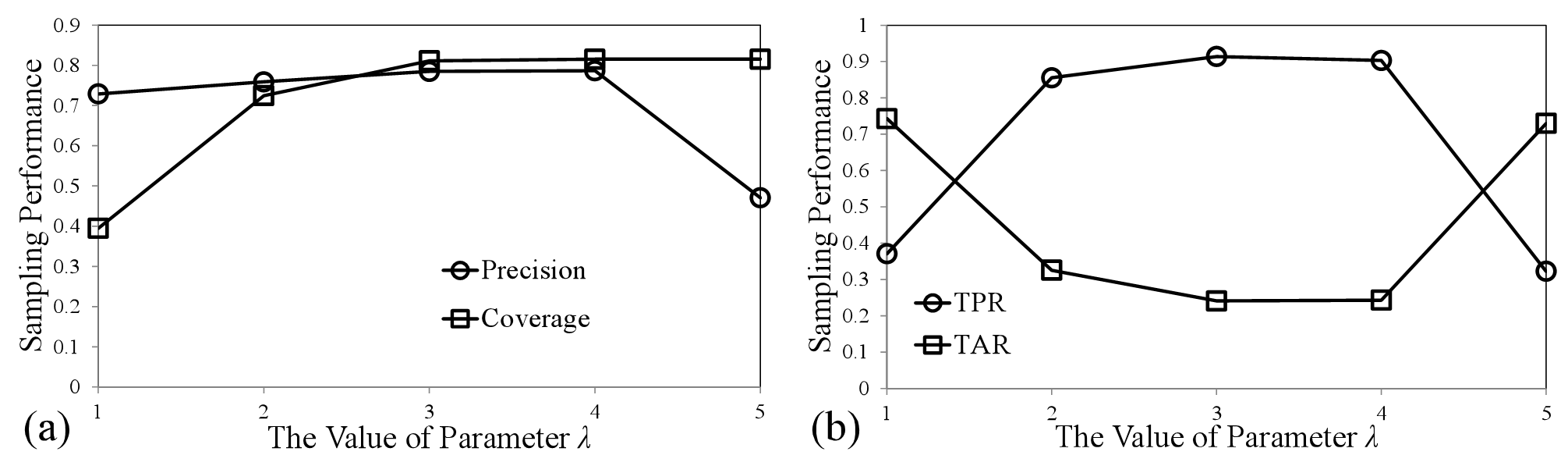}
	\caption{The tuning of parameter $\lambda$: (a) the performance of Precision and Coverage; (b) the performance of TPR and TAR.}
	\label{fig:8}
	\vspace{-0.2cm}
\end{figure}

The experiment results are shown in Figure \ref{fig:8}. First, with the increase of $\lambda$, there is a significant improvement in Coverage, reaching a plateau at $\lambda$ = 4. This is because the sampled technique subgraphs naturally become larger as the increase of $\lambda$, leading to a wider coverage. On the opposite, when $\lambda$ is too large, noisy NOIs that are incorrectly detected or belong to different APT technique instances might be merged into a single technique subgraph, leading to a cliff-like decline in Precision. Second, when $\lambda$ = 1 and $\lambda$ = 5, there is an extremely poor performance on TPR and TAR. When $\lambda$ is too small, it is difficult to find technique subgraphs with Coverage $>$ 0.8. When $\lambda$ is too large, it is difficult to guarantee the Precision of the sampled technique subgraphs. Both conditions would lead to fewer correct sampled technique subgraphs. The highest TPR and lowest TAR are both achieved at $\lambda$ = 3. Therefore, we set $\lambda$ = 3 in the following experiment.

\subsection{Experiment 3: Ablation Experiment}
In the first experiment, we try to evaluate the NOI detection model by comparing it with two state-of-the-art model as follows.

(1) StreamSpot \cite{r41}: It is a clustering-based anomaly detector. It abstracts the provenance graphs into a vector based on the relative frequency of local substructures, and detects abnormal provenance graphs as outliers based on a clustering algorithm. In order to detect abnormal nodes, we use the second-order local graph of a node as input.

(2) ThreaTrace \cite{r13}: It is a prediction-based anomaly detector. It learns every benign system entity’s representation in a provenance graph, and detects abnormal nodes based on the deviation from the predicted node type and its actual type.

The experiment results are shown in Table 6. First, TREC and ThreaTrace significantly outperform StreamSpot. It is because that TREC and ThreaTrace apply GNNs to learn the representation of nodes in a self-supervised way, while StreamSpot sketches the provenance graphs in an unsupervised way. The supervised signals could better optimize the representation model. Second, ThreaTrace has a slightly better performance than that of TREC in terms of Precision and FPR. It is because that ThreaTrace has designed a hierarchical combination of multiple classification models to reduce false alarms. However, the NOI detection model here is used to support the technique subgraph sampling task, so we believe Recall is more important than Precision. A high Recall could guarantee a sampled technique subgraph covers the complete malicious operations of an APT technique instance. Even if a high Precision could filter out most benign nodes, the DFS graph searching of the technique subgraph sampling algorithm would also append a large number of benign nodes to the technique subgraphs.

\begin{table}[t]
	\renewcommand{\arraystretch}{1.4}
	\caption{The comparison with state-of-the-art rule based methods.}
	\centering
	\label{tab:6} 
	\resizebox{0.48\textwidth}{!}{
		\begin{tabular}{ccccccc}
			\hline
			& Accuracy & Precision & Recall & F1 Score & FPR & AUC\\
			\hline
			StreamSpot & 0.856 & 0.733 & 0.741 & 0.737 & 0.251 & 0.743\\ 
			\hline
			ThreaTrace & 0.941 & 0.994 & 0.798 & 0.885 & 0.015 & 0.891\\ 
			\hline
			TREC & 0.943 & 0.755 & 0.912 & 0.834 & 0.067 & 0.954\\ 
			\hline
		\end{tabular}
	}
\end{table}
In the second experiment, we try to evaluate the technique subgraph representation learning model. Specifically, we test the recognition performance of the combinations of different meta-paths, i.e., $MPC_1$ to $MPC_6$ defined as follows.

(1)	$MPC_1$: It includes the meta-paths that only contain process (i.e., $MP_1$)

(2)	$MPC_2$: It includes the meta-paths that only contain process and file (i.e., $MP_1$ and $MP_2$).

(3)	$MPC_3$: It includes the meta-paths that only contain process, file, and socket (i.e., $MP_1$, $MP_2$, and $MP_4$).

(4)	$MPC_4$: It includes the meta-paths that only contain process, file, and registry (i.e., $MP_1$, $MP_2$, and $MP_3$).

(5)	$MPC_5$: It includes the meta-paths that only contain process, registry, and socket (i.e., $MP_1$, $MP_3$, and $MP_4$).

(6)	$MPC_6$: It includes all meta-paths (i.e., $MP_1$, $MP_2$, $MP_3$, and $MP_4$).

The experimental results are shown in Table \ref{tab:7}. First, $MPC_6$ exhibits the best performance, while $MPC_1$ performs the worst. It suggests that all meta-paths contribute to the subgraph representation, enabling the technique subgraph embeddings to retain more contextual information. This result also underscores the complexity of APT attacks. Second, $MPC_5$ outperforms $MPC_4$, and $MPC_4$ outperforms $MPC_3$. It suggests that “registry” is superior to “socket” and “socket” is superior to “file” in terms of providing effective information for APT tactic / technique recognition.

\begin{table}[t]
	\renewcommand{\arraystretch}{1.4}
	\caption{The performance of different meta-path combinations.}
	\centering
	\label{tab:7} 
	\resizebox{0.48\textwidth}{!}{
		\begin{tabular}{c|ccc|ccc}
			\hline
			& \multicolumn{3}{c|}{\textit{True\_Graph}} & \multicolumn{3}{c}{\textit{Sampled\_Graph}} \\
			\makecell[l]{Meta-path \\Combination} & TacACC & TechACC & SubACC & TacACC & TechACC & SubACC \\ 
			\hline
			$MPC_1$ & 0.799 & 0.745 & 0.566 & 0.578 & 0.272 & 0.123 \\
			$MPC_2$ & 0.818 & 0.653 & 0.684 & 0.556 & 0.318 & 0.260 \\
			$MPC_3$ & 0.799 & 0.758 & 0.566 & 0.674 & 0.381 & 0.291 \\
			$MPC_4$ & 0.847 & 0.796 & 0.742 & 0.727 & 0.646 & 0.491 \\
			$MPC_5$ & 0.866 & 0.771 & 0.754 & 0.642 & 0.670 & 0.613 \\
			$MPC_6$ & \textbf{0.875} & \textbf{0.797} & \textbf{0.766} & \textbf{0.813} & \textbf{0.70}3 & \textbf{0.672} \\
			\hline
		\end{tabular}
	}
\end{table}

In the third experiment, we try to evaluate the impact of the technique subgraph sampling on the final recognition performance. The experiment results are shown in Table \ref{tab:8}. First, it can be found that the performance of \textit{Sampled\_Graph} is significantly superior to \textit{Raw\_Graph} (especially TechACC and SubACC). This is because the original provenance graph samples without sampling are usually significantly larger than technique subgraphs. The vast majority of the nodes of an original provenance graph sample are noises (i.e., benign nodes). When computing the similarity between two samples, even the attention mechanism would assign higher weights to NOIs, the vast majority of benign nodes would still greatly disturb the similarity matching on malicious parts. Second, there is still a slight performance gap of APT technique recognition between \textit{True\_Graph} and \textit{Sampled\_Graph}. By analysing the experiment results, we find that the performance gap primarily stems from the imprecise technique subgraph sampling results.

\subsection{Experiment 4: Overhead Experiment}
\begin{figure}[h]
	\centering
	\includegraphics[width=0.48\textwidth]{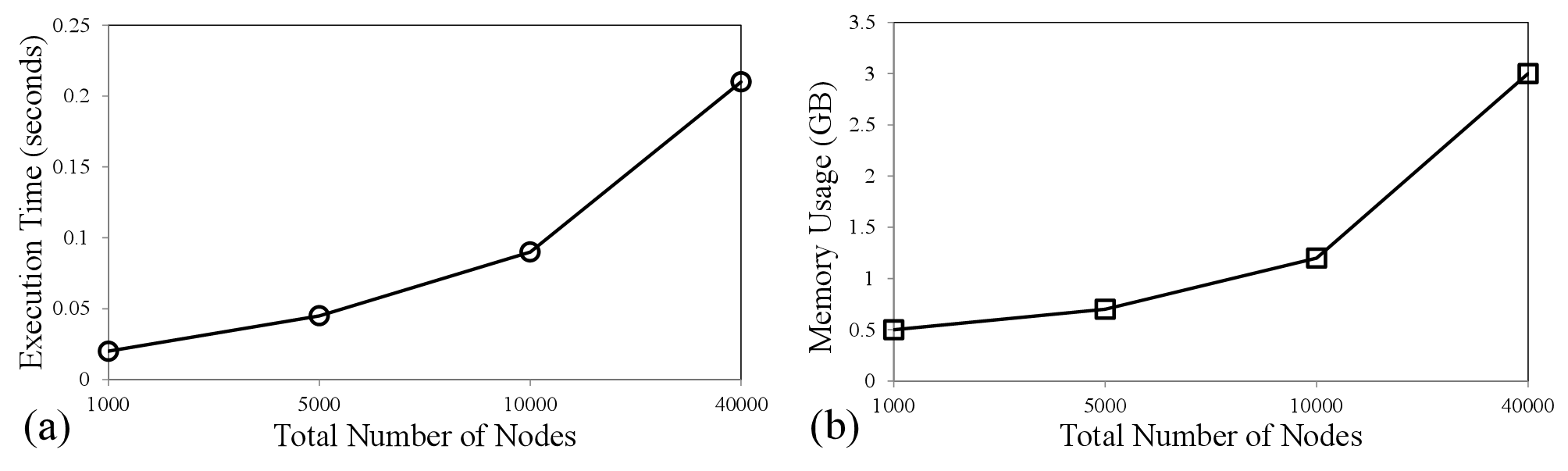}
	\caption{The impact of number of nodes on the computation overhead: (a) execution time (seconds); (b) memory usage (GB).}
	\label{fig:9}
\end{figure}

The main computational cost of TREC lies in the technique subgraph representation (Section 3.2) and matching (Section 3.3). Therefore, we monitor the execution time and memory usage of matching a pair of technique subgraphs with different number of nodes. Here, the matching operation includes reading two technique subgraphs into memory, generating embeddings for the two technique subgraphs, and calculating the similarity of them.
\begin{table}[h]
	\renewcommand{\arraystretch}{1.4}
	\caption{The performance of different technique subgraph inputs.}
	\centering
	\label{tab:8} 
	\resizebox{0.4\textwidth}{!}{
		\begin{tabular}{cccc}
			\hline
			& TacACC & TechACC & SubACC\\
			\hline
			\textit{Raw\_Graph} & 0.795 & 0.352 & 0.258 \\
			\textit{Sampled\_Graph} & 0.813 & 0.703 & 0.672 \\
			\textit{True\_Graph} & \textbf{0.875} & \textbf{0.797} & \textbf{0.766} \\
			\hline
		\end{tabular}
	}
\end{table}
The experiment results are shown in Figure \ref{fig:9}, where \textit{Num\_Nodes} stands for the total number of nodes in the two technique subgraphs. As \textit{Num\_Nodes} increases, both execution time and memory usage exhibit a clear growth. Note that the execution time and the memory usage do not increase linearly, because there are some computations (e.g., file reading, random walk in different graph structures, etc.) that are not linearly correlated to the number of nodes. In our dataset, the average number of nodes of technique subgraphs is 909, and there are 26 APT techniques and 45 APT sub-techniques, so the execution time of a complete matching operation for a technique subgraph is approximately 0.5 seconds for APT technique recognition and 0.9 seconds for APT sub-technique recognition.

\subsection{Discussion and Limitation}
\textbf{The prospect of TREC}: The main obstacle to applying deep learning techniques for APT detection is lack of malicious training samples. The experiment results show that TREC can achieve satisfactory performance on APT tactic recognition and fairly good performance on APT technique recognition and APT sub-technique recognition based on only several training samples for each APT technique / sub-technique. Given the results of this initial work on learning based APT tactic / technique recognition, we will try to further expand the experiment scale by considering more APT techniques and more diversified environments.

\textbf{The application of TREC}: The existing APT detection models (e.g., ThreaTrace \cite{r13}, APT-KGL \cite{r29}, etc.) and APT tracing systems (e.g., Nodoze \cite{r27}, DEPIMPACT \cite{r39}) are disjointed and studied separately. Most existing APT tracing systems assume that the malicious system entities are known in advance and have not considered the errors produced by APT detection models. TREC can act as an intermediate stage between APT detection and APT tracing. Specifically, APT detection models take low-level provenance data as input and output malicious system entities. TREC takes malicious system entities as input and output APT technique instances, which can compress alerts that may contain false alarms, hide detection errors, and provide semantic clues. Finally, APT tracing systems take APT technique instances as input and output the complete chain of the APT attack campaign.

The limitation of TREC: Compared with the extremely high accuracy of existing APT detection methods, there is still a certain gap in the accuracy of APT technique recognition of TREC. The reason is that the APT technique recognition task is far more difficult than the APT detection task. First, the APT technique recognition task has a significant larger label space, which further exacerbates the sample scarcity problem. Second, there are usually significant differences between benign and malicious samples, but the deviations between samples from different APT technique can be subtle. Some attacks with different APT techniques apply similar malicious operations. Aiming at this limitation, we will incorporate more data sources (e.g., network traffic, API call) and design more features (e.g., semantic features from file path, registry key, command line, and URL) to enhance the discriminative power of TREC.

\section{Related work}
\subsection{Provenance Based APT Detection}
According to the previous researches, provenance based analysis of audit logs is the best approach for the APT detection task\cite{r23,r24}. It converts the audit logs into a provenance graph, which can capture the temporal and causal relations between different system entities. Therefore, the long-term ongoing APT attacks can be detected and traced by reasoning on the paths on the provenance graph. Currently, the predominant strategy for provenance based analysis is rule-based. For example, PrioTracker\cite{r25} enables timely APT detection by computing the rareness scores to prioritize abnormal system events based on predefined rules. SLEUTH\cite{r7} and Morse\cite{r26} apply tag propagation rules to reconstruct APT attack scenarios through millions of system entities. NoDoze\cite{r27} uses contextual and historical information to assign anomaly scores to alerts based on score propagation rules. CONAN\cite{r9} proposes a state-based framework to consume system events as streams and represent each system entity in an FSA (Finite State Automata) like structure, in which the states are inferred through predefined rules. The advantages of the rule-based provenance analysis approaches are efficient and easy to deploy. However, they have low generalization ability and cannot detect new APT attacks.

Aiming at the limitations of the rule-based provenance analysis approaches, recent learning-based approaches, which could train models upon a large number of samples in an automatic way, start to gain more attention. Since provenance data can be inherently represented as graphs, graph based learning techniques are mostly exploited. According to the detected target, the existing work can be categorized into two types, i.e., node-level detection and subgraph-level detection.

For example of node-level detection, ThreaTrace\cite{r13} is an anomaly based detector that detects APT attacks at node level without prior knowledge of attack patterns by applying GraphSAGE model to learn the role of benign system entities in a provenance graph. ShadeWatcher\cite{r28} maps security concepts of system entity interactions to recommendation concepts of user-item interactions, and applies GNN based recommendation models to identify APT attacks. APT-KGL\cite{r29} learns a semantic vector representation for each system entity in a provenance graph based on a heterogeneous graph embedding model, and then detects malicious nodes based on a node classification model. Node-level detectors can provide fine-grained detection results, but the detection results cannot provide contextual knowledge (e.g., the relations between malicious nodes) that is valuable to understand the APT attack campaign.

For example of subgraph-level detection, ProvDetector\cite{r30} and ATLAS\cite{r31} identify possibly malicious nodes to form a path in the provenance graph, and then combine NLP (Natural Language Processing) and machine learning techniques to detect malicious paths. UNICORN\cite{r32} slices the whole provenance graph into evolving subgraphs by a sliding time window, and detects abnormal subgraphs by applying graph sketching to encode each subgraph into a fixed-size vector. Prov-Gem\cite{r33} proposes multi-embedding to capture varied contexts of nodes and aggregates the node embeddings to detect APT attacks at graph-level. ProGrapher\cite{r34} extracts temporal ordered subgraphs and performs detection on the subgraphs by combining graph embedding and sequence-based learning techniques. Subgraph-level detectors can detect APT attacks from a larger perspective, but the existing works have the following limitations. First, the subgraphs are roughly segmented (e.g., by time window, by random walk), and thus the malicious system entities in the subgraphs are not necessarily correlated. Second, most existing works perform a binary classification on the subgraphs to detect whether an attack behavior occurs or not without the detailed understanding or explanations of them.

\subsection{APT Tactic / Technique Recognition}
According to the Kill-Chain model\cite{r5} and AT\&CK knowledge base\cite{r6}, an APT attack campaign is usually composed of multiple attack steps over a long period of time. The above discussed APT detection methods can only detect a single attack step or even a single malicious activity in an attack step without the understanding of the big picture of the whole attack campaign. Since each attack step corresponds to an APT tactic / technique\cite{r35}, mapping each malicious node / edge or subgraph to an APT tactic / technique is obviously helpful to reconstruct the attack chain of an APT attack campaign.

Most existing works recognize APT tactics / techniques by designing a set of rules. For example, HOLMES\cite{r36} builds an intermediate layer between low-level provenance data and high-level Kill-Chain model, where the intermediate layer is defined as TTPs (Tactics, Techniques, and Procedures) in the ATT\&CK Matrix. The mapping between low-level provenance data and intermediate layer is achieved based on a set of TTP rules. RapSheet\cite{r37} also performs rule matching on provenance data to identify system events that match the APT tactics defined in the ATT\&CK Matrix. APTShield\cite{r38} defines a variety of labels for processes and files, and defines a set of transfer rules to propagate the labels according to the APT attack stages. Since APT attacks are complex and constantly evolving, static rules can be easily confused by slight deviation of attack patterns and cannot well cover the diverse APT tactics / techniques.

\section{Conclusion and Future Work}
In this paper, we propose TREC, a learning based system leveraging provenance data to recognize APT tactics / techniques. TREC employs a novel combination of techniques in anomaly detection, graph sampling, graph embedding, and distance metric learning, to detect and recognize APT tactic / technique instances from raw provenance graphs with very limited training samples. We evaluate TREC on a customized dataset collected by our team. As an initial work on learning based APT tactic / technique recognition, the experiment results show that TREC significantly outperforms state-of-the-art rule based systems on APT tactic recognition and has a fairly good performance on APT technique recognition. In practice, TREC can act as an intermediate stage between APT detection and APT tracing. TREC can arrange the fine-grained detection results of APT detection module, assign high-level semantics to these results, and then provide to the APT tracing module.

This work can be extended from the following directions in the future. First, most APT attack knowledge is shared by documents (e.g., CTIs) or rules (e.g., blacklists, IoCs). Therefore, we will try to extract attack knowledge and integrate it with machine learning methods for more robust APT tactic / technique recognition. Second, APT is a long-range attack campaign, and thus how to leverage the APT tactic / technique recognition model proposed in this paper to capture the big picture of the APT attack campaign will also be a promising research topic.

\begin{acks}
This work was supported by the National Natural Science Foundation of China (Nos. 62372410, U22B2028, 62072406) and the Zhejiang Provincial Natural Science Foundation of China (Nos. LZ23F020011, LDQ23F020001)
\end{acks}

\bibliographystyle{ACM-Reference-Format}
\bibliography{sample-base}

\newpage

\newpage
\appendix
\section{List of Simulated APT Tactics / Techniques}
The detailed list of APT tactics / techniques involved in our dataset is shown in Table \ref{tab:9}, where “\# Scripts” stands for the number of simulation scripts for the very APT tactic. The correspondence between APT tactics, APT techniques, and APT sub-techniques is based on ATT\&CK, and there exists APT techniques belonging to multiple APT tactics.

\section{Details of Re-implementation}
The re-implementation of the two rule based APT tactic / technique recognition systems (i.e., HOLMES and APTShield) is a challenging task. First, some rules are designed based on customized blacklists (e.g., the list of “Command\_Line\_Utilities” in HOLMES, the list of sensitive files in APTShield), which are not disclosed by the authors. Second, our dataset is collected from Windows system, but some rules are designed upon other operation systems. 

We use two strategies to address the two challenges. First, we create a customized blacklist for our dataset by extracting IoCs (Indicators of Compromise) from open CTIs (Cyber Threat Intelligences). The sources of CTIs and extraction methods are summarized in Table \ref{tab:10}. Second, we expanded the rules to adapt to Windows system (e.g., considering the Registry information).

According to the two strategies, we elaborate the re-implementation process of HOLMES and APTShield as follows.

\subsection{HOLMES}
The rules in HOLMES are defined based on the seven APT tactics in Kill-Chain, so we align the APT tactics in ATT\&CK with those in Kill-Chain according to Table \ref{tab:11}, and then we compare TREC and HOLMES under the Kill-Chain system.

\subsection{APTShield}
The rules in APTShield are defined based on Linux system and cannot be directly applied to Windows system. Therefore, we first updated the system entity state definitions by modifying the state definitions of process and file (in Table \ref{tab:12}) and adding new state definitions of registry (in Table \ref{tab:13}). Second, we updated the state transfer rules to adapt to the new state definitions (in Table \ref{tab:14}).\

\begin{table}[b]
	\renewcommand{\arraystretch}{1.4}
	\caption{The alignment between APT tactics in Kill-Chain and ATT\&CK.}
	\centering
	\label{tab:11} 
	\resizebox{0.48\textwidth}{!}{
		\begin{tabular}{cc}
			\hline
			\textbf{APT Tactics in Kill-Chain} &\textbf{ APT Tactics in ATT\&CK} \\
			\hline
			Initial Compromise & \makecell[l]{Initial Access} \\ [0.5em]
			Establish Foothold & \makecell[l]{Execution, Persistence, Command and Control, \\Defense Evasion} \\ [1em]
			Privilege Escalation & \makecell[l]{Privilege Escalation, Defense Evasion} \\[0.5em]
			Internal Recon & \makecell[l]{Discovery, Collection, Defense Evasion} \\[0.5em]
			Move Laterally & \makecell[l]{Lateral Movement, Defense Evasion} \\[0.5em]
			Complete Mission & \makecell[l]{Collection, Command and Control, Impact} \\[0.5em]
			Cleanup Tracks & \makecell[l]{Impact, Defense Evasion} \\[0.5em]
			\hline
		\end{tabular}
	}
\end{table}

\begin{table*}[!htbp]
	\renewcommand{\arraystretch}{1.4}
	\caption{The modified state definitions of process.}
	\centering
	\label{tab:12} 
	\resizebox{0.9\textwidth}{!}{
			\begin{tabular}{|c|c|c|c|}
					\hline
					\textbf{\makecell[l]{Entity Type}} &\textbf{\makecell[l]{State Code}} & \textbf{Description} & \textbf{\makecell[l]{State Type}}\\
					\hline
					\multirow{17}{*}{Process} & PS1 & \makecell[l]{The process has a network connection. }& \multirow{9}{*}{Stage}\\
					\cline{2-3}
					& PS2 & \makecell[l]{The process accessed a high-value file.} & \\
					\cline{2-3}
					& PS3 & \makecell[l]{The process contains network data.} & \\
					\cline{2-3}
					& PS4 & \makecell[l]{The process reads or loads files from the external network.} & \\
					\cline{2-3}
					& PS5 & \makecell[l]{The process reads sensitive system information.} & \\
					\cline{2-3}
					& PS6 & \makecell[l]{The process reads account related information.} & \\
					\cline{2-4}
					& PB1 & \makecell[l]{The process executes files from the network.} & \multirow{10}{*}{Behavior} \\
					\cline{2-3}
					& PB2 & \makecell[l]{The process executes sensitiive files.} & \\
					\cline{2-3}
					& PB3 & \makecell[l]{The process executes sensitive commands.}& \\
					\cline{2-3}
					& PB4 & \makecell[l]{The process modified the security control policy.} & \\
					\cline{2-3}
					& PB5 & \makecell[l]{The process modified the scheduled task policy.} & \\
					\cline{2-3}
					& PB6 & \makecell[l]{The process modified the permission control policy.} & \\
					\cline{2-3}
					& PB7 & \makecell[l]{The process reads high-value information.} & \\
					\hline
					\multirow{9}{*}{File} & FU1 & \makecell[l]{It is an uploaded file.} & \multirow{3}{*}{Untrusted}\\
					\cline{2-3}
					& FU2 & \makecell[l]{The file contains data from the Web.} & \\
					\cline{2-3}
					& FU3 & \makecell[l]{The file does not exist.} & \\
					\cline{2-4}
					& FH1 & \makecell[l]{The file contains control scheduled tasks.} & \multirow{6}{*}{High Value}\\
					\cline{2-3}
					& FH2 & \makecell[l]{The file contains control user permissions.} & \\
					\cline{2-3}
					& FH3 & \makecell[l]{The file contains sensitive user information.} &\\
					\cline{2-3}
					& FH4 & \makecell[l]{The file contains security control policy.} &\\
					\cline{2-3}
					& FH5 & \makecell[l]{The file was written by a process that read sensitive information.}&\\
					\cline{2-3}
					& FH6 & \makecell[l]{The file contains sensitive system information.} & \\
					\hline
				\end{tabular}
		}
\end{table*}


 \begin{table*}[t]
		 \renewcommand{\arraystretch}{1.4}
		 \caption{The modified state definitions of file.}
		 \centering
		 \label{tab:13} 
		 \resizebox{0.9\textwidth}{!}{
				   \begin{tabular}{|c|c|c|c|}
						     \hline
						     \textbf{\makecell[l]{Entity  Type}} &\textbf{\makecell[l]{State Code}} & \textbf{Description} & \textbf{\makecell[l]{State Type}}\\
						     \hline
						     \multirow{9}{*}{File} & FU1 & \makecell[l]{It is an uploaded file.} & \multirow{3}{*}{Untrusted}\\
						     \cline{2-3}
						      & FU2 & \makecell[l]{The file contains data from the Web.} & \\
						      \cline{2-3}
						      & FU3 & \makecell[l]{The file does not exist.} & \\
						      \cline{2-4}
						      & FH1 & \makecell[l]{The file contains control scheduled tasks.} & \multirow{6}{*}{High Value}\\
						      \cline{2-3}
						      & FH2 & \makecell[l]{The file contains control user permissions.} & \\
						      \cline{2-3}
						      & FH3 & \makecell[l]{The file contains sensitive user information.} &\\
						      \cline{2-3}
						      & FH4 & \makecell[l]{The file contains security control policy.} &\\
						      \cline{2-3}
						      & FH5 & \makecell[l]{The file was written by a process that read sensitive information.}&\\
						      \cline{2-3}
						      & FH6 & \makecell[l]{The file contains sensitive system information.} & \\
						      \hline
						   \end{tabular}
				   }
		 \end{table*}

\begin{table*}[!b]
	\renewcommand{\arraystretch}{1.4}
	\caption{The list of APT tactics / techniques in our dataset.}
	\centering
	\label{tab:9} 
		\begin{tabular}{cccc}
			\hline
			APT Tactic & APT Technique & APT Sub-Technique & \# Scripts \\
			\hline
			Initial Access & \makecell[l]{T1566} &  \makecell[l]{T1566.001} & 2 \\[0.5em]
			Credential Access & \makecell[l]{T1558, T1552, T1003, T1555, T1110, \\T1556} & \makecell[l]{T1558.001, T1552.004, T1558.004, T1003.001, T1003.002,\\ T1003.003, T1552.001, T1552.002, T1555.003, T1552.006,\\ T1110.003, T1003.006, T1556.002} & 92 \\[1.5em]
			Defense Evasion & \makecell[l]{T1562, T1218, T1070, T1548, T1222, \\T1556, T1134, T1036} & \makecell[l]{T1562.006, T1562.004, T1562.001, T1218.007, T1218.005, \\T1218.011, T1218.004, T1070.004, T1548.002, T1222.001,\\ T1562.002, T1070.005, T1218.010, T1556.002, T1134.004, \\T1218.001, T1036.003} & 211 \\[2em]
			Persistence & \makecell[l]{T1547, T1137, T1543, T1556, T1053} & \makecell[l]{T1547.001, T1137.006, T1547.004, T1543.003, T1556.002,\\ T1053.005} & 64 \\[1em]
			Privilege Escalation & \makecell[l]{T1547, T1548, T1543, T1134, T1053} & \makecell[l]{T1547.001, T1548.002, T1547.004, T1543.003, T1134.004, \\T1053.005} & 85 \\[1em]
			Discovery & \makecell[l]{T1087, T1069, T1614} & \makecell[l]{T1087.002, T1069.002, T1614.001, T1069.001} & 46 \\[0.5em]
			Execution & \makecell[l]{T1204, T1569, T1059, T1053} &\makecell[l]{ T1204.002, T1569.002, T1059.001, T1059.003, T1053.005} & 60 \\[0.5em]
			Exfiltration & \makecell[l]{T1048} & \makecell[l]{T1048.003} & 9 \\[0.5em]
			Impact & \makecell[l]{T1491} & \makecell[l]{T1491.001} & 2 \\[0.5em]
			\hline
		\end{tabular}
\end{table*}

\begin{table*}[!htbp]
	\renewcommand{\arraystretch}{1.4}
	\caption{The sources and extraction methods of CTIs.}
	\centering
	\label{tab:10} 
	\resizebox{0.95\textwidth}{!}{
		\begin{tabular}{|c|l|c|c|}
			\hline
			\textbf{CTI Source} &\textbf{\makecell[c]{CTI Example}} & \textbf{Extraction Method} & \textbf{IoC Example}\\
			\hline
			ATT\&CK & 
			\makecell[l]{Many OS utilities may provide information about local device\\ drivers, such as driverquery.exe and the EnumDeviceDrivers() \\API function on Windows. Information about device drivers \\(as well as associated services, i.e., System Service Discovery)\\ may also be available in the Registry.} &
			\makecell[c]{Regular \\Pattern Match} & \makecell[l]{driverquery.exe \\ EnumDeviceDrivers()} \\[3em]
			\hline
			Red Canary &
			\makecell[l]{
				...\\
				Copy-Item Copy-Item "\$env:Temp\textbackslash NPPSPY.dll" \\-Destination "C:\textbackslash Windows\textbackslash System32" \\
				\$path = Get-ItemProperty -Path \\"HKLM:\textbackslash SYSTEM\textbackslash CurrentControlSet\textbackslash Control\textbackslash NetworkProvider\textbackslash Order"\\ -Name PROVIDERORDER\\
				\$UpdatedValue = \$Path.PROVIDERORDER + ",NPPSpy"\\
				Set-ItemProperty -Path \$Path.PSPath -Name "PROVIDERORDER" \\-Value \$UpdatedValue \\
				...
			}
			& \makecell[c]{Regular \\Pattern Match} &
			\makecell[l]{Temp\textbackslash NPPSPY.dl\\
				C:\textbackslash Windows\textbackslash System32\\
				HKLM:\textbackslash SYSTEM\textbackslash Cu\\rrentControlSet\textbackslash Control\textbackslash \\ NetworkProvider\textbackslash Order
			} \\
			\hline
			Sigma-Rule &
			\begin{tabular}{ll}
				\makecell[l]{
					...\\
					tags:\\
					\hspace{1em}- attack.impact\\
					\hspace{1em}- attack.t1490\\
					...\\
				}
				&
				\makecell[l]{
					...\\
					detection:\\
					\hspace{1em}selection:\\
					\hspace{2em}Image|endswith:\\
					\hspace{3em}- '\textbackslash cmd.exe'\\
					\hspace{3em}- '\textbackslash powershell.exe'\\
					\hspace{3em}- '\textbackslash pwsh.exe'\\
					\hspace{3em}- '\textbackslash wt.exe'\\
					\hspace{3em}- '\textbackslash rundll32.exe'\\
					\hspace{3em}- '\textbackslash regsvr32.exe'\\
					...
				}
			\end{tabular}
				% ...
			& Yaml Parsing 
			& \makecell[l]{
				\textbackslash cmd.exe\\
				\textbackslash powershell.exe\\
				\textbackslash pwsh.exe\\
				\textbackslash wt.exe\\
				\textbackslash rundll32.exe\\
				\textbackslash regsvr32.exe 
			}\\
			\hline
		\end{tabular}
	}
\end{table*}

\begin{table*}[h]
	\renewcommand{\arraystretch}{1.4}
	\caption{ The updated state transfer rules.}
	\centering
	\label{tab:14} 
	\resizebox{0.9\textwidth}{!}{
		\begin{tabular}{|c|c|c|c|c|}
			\hline
			\textbf{APT Tactic} &\textbf{State A} & \textbf{State B} & \textbf{Event} & Description\\
			\hline
			\multirow{4}{*}{Initial Access} & PS1 & FU2 & Write & A process with a network connection writes the file. \\
			\cline{2-5}
			& PS3 & FU2 & Read & A process reads a file containing network data. \\
			\cline{2-5}
			& PS3 & FU2 & Write & A process that has accessed network data writes a file. \\
			\cline{2-5}
			& PS4 & FU1 & Read/Image Load & A process loads or reads files uploaded by the user. \\
			\hline
			\multirow{5}{*}{Execution} & PB1 & FU2 & Execute & A network file is executed. \\
			\cline{2-5}
			& PB1 & FU2 & Image Load & A network file is loaded. \\
			\cline{2-5}
			& PB1 & FU2 & Write & A process that has executed the network file writes a file. \\
			\cline{2-5}
			& PS5 & RU2 & Create Registry & A process that has read sensitive information modifies the registry key. \\
			\cline{2-5}
			& PS5 & RU2 & Read Registry & A process reads a registry key that may contain sensitive information. \\
			\hline
			\multirow{3}{*}{Persistence} & PB6 & FH1 & Write & A process writes to a file with controlled scheduled tasks. \\
			\cline{2-5}
			& PB6 & RH1 & Modify Registry & A process modifies the registry key that controls scheduled tasks. \\
			\cline{2-5}
			& PS4 & RU3 & Create Registry & A process that loads or executes suspicious files creates a registry key. \\
			\hline
			\multirow{3}{*}{Privilege Escalation} & PB7 & FH2 & Write & A process writes to a file that controls permissions. \\
			\cline{2-5}
			& PB7 & RH1 & Modify Registry & A process modifies the registry key that controls permissions. \\
			\hline
			\multirow{2}{*}{Credential Access} & PS6 & FH3 & Read & A process writes to files with sensitive information. \\
			\cline{2-5}
			& PS6 & RH3 & Read Registry & A process reads a registry key with sensitive information. \\
			\hline
			\multirow{2}{*}{Defense Evasion} & PB5 & FH4 & Write & A process writes to a file with security control information. \\
			\cline{2-5}
			& PB5 & RH4 & Modify Registry & A process modifies the registry key with security control information. \\
			\hline
			\multirow{2}{*}{Discovery} & PS5 & FH6 & Read & A process reads system-sensitive files. \\
			\cline{2-5}
			& PS5 & RH6 & Read Registry & A process reads system-sensitive registry key. \\
			\hline
			\multirow{2}{*}{Data Exfiltration} & PB5-7\textbar PS2, 5-6 & FH5 & Write & A process writes high-value data to files. \\
			\cline{2-5}
			& PB8 & FH5 & Read & A process reads high-value data files. \\
			\hline
		\end{tabular}
	}
\end{table*}

\end{document}